**Title:** Enumerating the climate impact of disequilibrium in critical mineral supply

**Authors:** Lucas Woodley[1], Chung Yi See[2], Peter Cook[3], Megan Yeo[2], Daniel S. Palmer[4], Laurena Huh[5], Seaver Wang[3], and Ashley Nunes[2,3,6]

[1] Faculty of Arts and Sciences
Harvard University
Cambridge, MA, 02138, USA

[2] Department of Economics
Harvard College
Cambridge, MA, 02138, USA

[3] Breakthrough Institute
Berkeley, CA 94704

[4] The Groton School
Groton, MA 01450

[5] Sloan School of Management
Massachusetts Institute of Technology
Cambridge, MA 02138

[6] Center for Labor and a Just Economy
Harvard Law School
Cambridge, MA, 02138, USA

**Corresponding Author:** Ashley Nunes, ashley@thebreakthrough.org

**Abstract**

Recently proposed tailpipe emissions standards aim to significant increases in electric vehicle (EV) sales in the United States. How achievable is this increase given potential constraints in EV mineral supply chains? Our work addresses this question. We estimate a model that reflects nation-of-origin sourcing rules, heterogeneity in the mineral intensity of predominant battery chemistries, and long-run grid decarbonization efforts. Our efforts yield five key findings. First, compliance with the EPA's proposed standard necessitates replacing at least 10.21 million new ICEVs with EVs between 2027 and 2032. Second, based on economically viable and geologically available mineral reserves, manufacturing sufficient EVs is plausible across most battery chemistries and could – subject to the chemistry leveraged – reduce up to 457.3 million total tons of $CO_2$e. Third, mineral production capacities of the US and its allies constrain battery production to a total of 5.09 million EV batteries between 2027 and 2032, well short of deployment requirements to meet EPA standards even if battery manufacturing is optimized to exclusively manufacture materials efficient NMC 811 batteries. Fourth, disequilibrium between mineral supply and demand results in at least 59.54 million tons of $CO_2$e in total lost lifecycle emissions benefits. Fifth, limited present-day production of battery-grade graphite and to a lesser extent, cobalt, constrain US electric vehicle battery pack manufacturing under strict nation-of-origin content sourcing rules. We demonstrate that should mineral supply bottlenecks persist, hybrid electric vehicles (HEVs) may offer equivalent lifecycle emissions benefits as EVs while relaxing mineral production demands, though this represents a tradeoff of longer-term momentum in electric vehicle deployment in favor of near-term CO2 reductions.

**Introduction**

Acceleration of the energy transition and realization of both national and regional climate commitments require urgent action on a global scale (1). These goals depend upon adoption of technologies that facilitate emissions reductions. However, energy systems powered by low-carbon technologies differ profoundly from current systems of fossil fuel trade and infrastructure. The manufacturing of solar photovoltaic plants, wind farms and electric vehicles (EVs) – technologies crucial to lowering emissions – generally requires considerable volumes of specialty minerals, with mineral intensity varying greatly across different technologies (2-4).

Meeting the mineral demands associated with electrifying the light-duty vehicle fleet warrants particular attention given the transportation sector's contribution to $CO_2$ emissions. Owing to an existing internal combustion engine vehicles' (ICEVs) dependence on fossil-fuels, cars, vans, and sport utility vehicles produce nearly half of all transportation-related greenhouse gas emissions, making these vehicles significant contributors to climate change (5,6). Electrification offers – by virtue of reduced dependence on fossil fuels – a lower well-to-wheels emissions profile, which can reduce overall emissions relative to the status quo (7,8).

However, raw material supply chain bottlenecks present potential obstacles for an efficient transition to EVs. Of the eight minerals used in EVs (excluding ferrous metals and aluminum), five minerals- cobalt, graphite, lithium, nickel, and rare earths - are not used to any significant degree in ICEVs. Moreover, relative to ICEVs, an EV requires twice the weight of copper and manganese – two additional key minerals (2). Higher mineral demands imposed by EVs (relative to ICEVs) and the envisioned prospect of widespread electrification as a pathway towards emissions reduction raise the important question: do mineral demands associated with electrifying the light-duty vehicle fleet exceed available supply? If so, by how much? And what are the emissions consequences of disequilibrium in critical minerals market?

Answers to these questions are timely, particularly for the United States where emissions from the largely ICE-powered light-duty vehicle fleet constitute a significant share of overall emissions. This share has grown over time and appears likely to continue as household motorization rates rise (9). The U.S. federal government has – since 2008 – sought to temper the emissions impact of ICEVs by incentivizing EV adoption (10). The most recent example of these efforts are tailpipe emissions standards recently proposed by the Environmental Protection Agency (EPA) (11). The standards necessitate – for vehicles sold after 2026 – realization of an industry-wide average target for the light-duty fleet of 82 grams/mile (g/mile) of $CO_2$ by 2032. Given current market conditions and the thermal inefficiency of ICEVs, achieving this target necessitates significantly increasing EV sales volume (12). How achievable is the requisite increase given constraints in EV mineral supply chains?

Our work addresses this question. To do so, we estimate a model that, 1) explores requisite EV sales volume scenarios that conform to the U.S. light-duty vehicle electrification targets set by the EPA, 2) assesses whether existing U.S. mineral supply chains can accommodate EV manufacturing levels needed to achieve the sales targets associated with these scenarios, and 3) quantifies the emissions impact of potential disequilibrium between mineral supply and EV-associated mineral demand for each scenario. Our model accommodates envisioned improvements in the emissions profile of alternative powertrains owing to technological and legislative efforts, most notably, Corporate Average Fuel Economy standards (CAFE) for light duty vehicles between 2027 and 2032 (13-15). We further consider EVs' potential to operate as substitutes rather than complements (16,17) and envisioned reductions in the carbon intensity of the electrical grid that may – owing to legislation like the 2022 Inflation Reduction Act – improve the emissions profile of EVs (18).

Three key characteristics define our approach. First, in estimating potential mineral disequilibrium, we recognize that the US is ill-suited to pursue full self-sufficiency in several key minerals – most notably cobalt, graphite, and manganese – that are necessary to produce lithium-ion batteries used in EVs (18). We concurrently acknowledge political concerns that some countries levy disproportional influence over key aspects of the automotive supply chain and may use raw material and manufacturing market power not only to limit supplies, but also to further attract and concentrate foreign investment and advanced manufacturing (2,19,20). Consequently, consistent with the intent of mineral sourcing provisions of the Inflation Reduction Act, our estimates consider minerals that are available either domestically or from countries with which the US has free trade and/or mutual defense agreements. Furthermore, our analysis distinguishes between limits in the US and partner countries' annual total rate of upstream mineral production (hereafter referred to as production) and total geologic mineral reserves across those same countries (hereafter referred to as reserves) (see Methods for details).

Second, our efforts consider heterogeneity in adoption of specific battery chemistries and the emissions intensity associated with these chemistries (21-23). Electrification policies do not – to our knowledge – prioritize one battery chemistry over another but rather emphasize specific EV penetration rates. Yet, consideration of the emissions intensity associated with extracting minerals specific to a particular battery chemistry is timely because it influences the magnitude of total emissions reductions EVs ultimately deliver. Put simply, EVs utilizing relatively carbon-intense battery chemistries likely offer a smaller emissions benefit – ceteris paribus - than chemistries with lower manufacturing-related emissions. To account for such factors, we leverage the 2022 Greenhouse Gases, Regulated Emissions, and Energy Use in Transportation (GREET) model to estimate emissions associated with adopting specific battery chemistries (24).

Third, we assess whether specific emissions reduction targets envisioned by the EPA can – given potential constraints in mineral supplies - be realized by deploying a wider combination of vehicle powertrains in the national fleet, namely hybrid electric vehicles (HEV). The manufacturing of HEV batteries requires fewer specialty raw materials, alleviating mineral supply constraints (25,26). Moreover, HEVs offer substantially lower emissions relative to ICEVs for a similar vehicle price (thereby offering greater affordability to consumers relative to current EVs) (27) while also enjoying relative market popularity compared to EVs (thereby affording more rapid widespread deployment) (28). HEVs' popularity has persisted despite the gradual withdrawal of HEV-specific procurement incentives first enacted in 2008 (29,30). Consequently, we also explore whether complementary deployment of HEVs can help drive near-term transportation sector emissions reductions while alleviating immediate raw material supply chain constraints confronting EVs.

Our efforts can help better inform public policies that target transportation-related emissions reductions in the face of potential mineral supply constraints on EV battery pack manufacturing. Furthermore, by scrutinizing geographic patterns associated with mineral supply constraints, our work can inform efforts to address economic and national security concerns related to possible mineral shortfalls. As countries like the United States accelerate their efforts to reduce emissions and deploy new low-carbon technologies, policymakers must create the underlying conditions for a new generation of technologies to achieve widespread adoption while maintaining reliable and affordable energy and mobility systems amidst real-world constraints (3). In the long term, EVs appear poised to dominate the future of clean transportation. In the medium term, however, the tension between ambitious policy targets with fixed timetables and the inertia facing supply chain expansion poses complex challenges. A better understanding of the linkages between raw material availability, battery pack chemistries, and the advantages and drawbacks of different low-emissions vehicle types improves assessments of different policy options' impacts, thereby promoting more effective public policy.

**Results and Discussion**

Electrification of the U.S. light-duty vehicle fleet carries the potential to reduce CO2 emissions, public health harm from air pollution, and national dependence on fossil fuels. These societal benefits have prompted the EPA to propose stringent emissions standards that de facto necessitate EV adoption. How achievable are the proposed standards given constraints in mineral supply chains?

We address this question by, 1) specifying requisite EV sales volume targets across three sales scenarios (low, medium, and high) that each conform to the electrification targets set by the EPA, 2) enumerating the extent to which these targets can be met by using a single battery chemistry (referred to as 'optimal chemistry') or combination thereof (referred to as 'market mix'), and 3) quantify the emissions impact of disequilibrium between mineral supply and demand. Realization of EPA prescribed sales volume targets are estimated using both mineral reserve and mineral production estimates (see Method for details). Where our model produces different estimates for each sales scenario, we present results for the medium sales scenario followed by the range across the low and high scenarios in parentheses.

Our key findings – summarized in Figure 1 - are as follows.

First, we find that by the year 2032 - given projected emission profiles of ICEVs and HEVs, - 37.82 percent of new light-duty vehicle sales must be EVs for auto manufacturers to ensure compliance with the EPA's tailpipe emissions proposal. This finding – which is lower than other projected estimates regarding requisite EV market share (12) – reflects the impact that improved fuel economy of non EVs (necessitated by the most recent CAFE standards update), have on requisite EV penetration rates. Given the interdependencies between fuel economy and tailpipe emissions, a fossil fuel powered light duty vehicle fleet with higher fuel economy is less polluting, which in turn requires lower requisite EV market share to comply with the EPA's tailpipe emissions standard. Nevertheless, requiring that 37.82 percent of new light-duty vehicle sales be EVs requires – consistent with the envisioned intent of the EPA proposal - a significant increase in EV sales relative to the present day. Assuming the overall size of the light duty vehicle fleet remains consistent with government projections, our model estimates that new EVs must displace 28.05 million new ICEVs between 2027 and 2032 (10.21 and 34.62 million in the low and high sales scenarios, respectively) to comply with the proposed rule (see Table 1).

Second, we find that from the vantage point of mineral reserves alone, supporting the requisite number of vehicles required for EPA compliance is plausible across all scenarios for five of the six battery chemistries investigated. That is, the total quantities of economically extractable minerals contained in the US and partner countries is theoretically sufficient to meet the required magnitude of EV deployment. Specifically, we find that for the deployment of EVs using solely NMC 523, NMC 622, NMC 811, NCA or LFP batteries, leveraging mineral reserves can support between 81.66 million and 989.27 million EVs. This well exceeds the 34.62 million EVs estimated for compliance in our high penetration scenario, reducing lifecycle emissions by up to 457.3 million tons of $CO_2e$. Reliance on LFP battery chemistry maximizes the number of EVs supported (989.27 million), followed by NCA (400.37 million), NMC 811 (201.80 million), NMC 523 (90.21 million) and NMC 622 (81.66 million). Moreover, leveraging a combination of LFP and NCA chemistries affords additional EV batteries to be manufactured; namely, available reserves can simultaneously produce 735.19 million LFP batteries and 400.37 million NCA batteries, thereby supporting a total of 1.14 billion EVs. Reliance on NMC 111 exclusively affords the fewest number of vehicles supported (47.71 million).

Although these results imply that vehicle manufacturers can fully satisfy EV demand using numerous potential major chemistries (31), access to geological mineral reserves depends in practice upon mineral production capacity. Whereas reserves refer to long term, cumulative economically viable supply, production rates reflect existing extraction and processing capacity. Consequently, in addition to solely considering geological reserves, planners must assess whether available mineral production capacity can enable realization of the EPA's electrification targets.

Our third finding is that, based on current mineral production from the US and its allies between 2027 and 2032, a maximum of 5.09 million EV batteries can be produced cumulatively, a figure that falls well below of the requisite number of EV batteries in even our lowest sales scenario (10.2 million) (see Table 1). Graphite is the key limiting mineral driving battery chemistry choice that maximizes potential EV deployment, as exclusive manufacturing of NMC 811 EV batteries supports no more than 5.09 million vehicles (Fig. 2a). This effect is sensitive to our input battery mineral intensity data, which assume 56.6 kg of graphite for a 75 kWh NMC 811 battery pack, with alternative chemistries such as NCA or LFP requiring even higher amounts of graphite (Fig. 2b). Specifically, we find limits of 4.70 and 2.98 million NCA and LFP batteries, respectively. Under these assumptions, graphite might potentially pose a challenge to envisioned market shifts towards NCA and LFP EVs by 2032. Moreover, were LFP battery packs to be increasingly favored owing to their cost advantage (32), we find that production capacity from the US and its allies' alone may currently support manufacturing of only 3.51 million LFP battery packs from 2027 – 2032.

What are the emissions consequences of being unable to fully meet the EPA's implied EV sales targets? Our fourth finding is that, assuming manufacturers maximize the quantity of available EVs in each given year by utilizing NMC 811 chemistries exclusively, the US light-duty vehicle fleet will contain 22.96 million (5.12 to 29.53 million) fewer EVs than the EPA targets. This shortfall is equivalent to 284.12 million tons $CO_2$e (59.54 million to 369.05 million tons $CO_2$e) in lost lifecycle emissions benefits. Meanwhile, if the EV fleet evolves using a mix of battery chemistries, the light-duty vehicle fleet may contain 24.54 million (6.70 to 31.11 million) fewer EVs than the EPA targets, which is analogous to 310.56 million tons $CO_2$e (81.11 million to 397.23 million tons $CO_2$e) in lost lifecycle emissions benefits.

*Resolution pathways*

Given the potential for mineral constraints to impede the effectiveness of the EPA's proposed rule, how can policymakers respond? We investigate two potential pathways.

The first entails increasing mineral production capacity to better meet the mineral demand requirements imposed by the EPA's emissions proposal. Given that this approach is currently the focus of ongoing discussions, most notably through domestic mine permitting reform (33-38), we enumerate specific production thresholds that warrant consideration to fully realize the EPA's electrification goals. Graphite is the primary constraining battery material, while cobalt would also pose an obstacle to the required magnitude of EV deployment in scenarios that rely more heavily on NMC and NCA chemistries.

Conversely, increased production of other key minerals (i.e., Aluminum, Copper, Lithium, Manganese, Nickel, and Phosphate) – absent increases in graphite and cobalt – does not increase the number of EV battery packs that can be manufactured. We therefore direct scrutiny towards the requisite increases to graphite and cobalt production.

Based on existing industry announcements for new natural graphite mines and synthetic graphite plants in the US, production of graphite by the US and partner countries could increase to 173,000 tons per year by 2026, while from 2027 to 2032, graphite production would further increase to 255,000 tons per year (33-37). However, achieving the required annual US EV sales of 5.71 million in 2032 will require up to 470,000 tons per year of available battery-grade graphite. This constitutes an 880 percent increase in production relative to the present day and exceeds the projected increases based on existing announcements. Cobalt production by the US and eligible partners would need to meet demand of up to 31,000 tons per year, or up to 42 percent more than current collective production rates, to support US light-duty vehicle electrification.

Beyond graphite and cobalt, mineral constraints become less acute. Between the US and partner countries, present-day mineral production for aluminum, copper, lithium, manganese, nickel, and phosphate would theoretically suffice to meet US vehicle electrification goals under all sales scenarios. Although, we note that lithium and nickel – which are the next most limiting raw materials – could pose potential challenges owing to their use outside the electric light-duty vehicle sector (heavy trucks, consumer electronics, utility-scale batteries, other electric mobility technologies).

To what degree would a universal increase in mineral production support additional EV deployment? To assess the sensitivity of battery pack manufacturing limits to changes in mineral availability, we incorporate an Added Supply Assumption in our model. Here, we assume that for each mineral, an additional amount of eligible mineral production becomes available for US EV manufacturing, with that amount equivalent to 20 percent of the national annual production of the world's leading supplier of each mineral. We find that the Added Supply Assumption dramatically alleviates graphite constraints, enabling cumulative 2027-2032 deployment of 23.13 million EVs if exclusively manufacturing NMC811 battery packs, or 15.95 million EVs for a market mix of battery chemistries. Relative to current mineral production, this represents a significant increase in deployable EVs using both an optimal chemistry (initially 5.09 million from 2027 – 2032 when using NMC 811 batteries) and a market mix (initially 3.51 million from 2027 – 2032). However, leveraging the Added Supply Assumption still produces an EV sales shortfall in the medium and high sales scenario (that require 28.05 million and 34.62 million EV sales respectively).

Additionally, automobile manufacturers could conceivably consider smaller EV battery packs to stretch mineral supplies further. Meeting EPA emissions standards requires implied nationwide deployment of at least 5.71 million EVs in the year 2032. At present-day rates of production (48,000 tons per year), graphite remains – we find – a constraining factor, allowing for just 8.4 kg of graphite per battery pack. This corresponds to a graphite intensity approximately 56 percent that of a typical 20 kWh NMC811 battery like those used in plug-in hybrids, therefore yielding a battery capacity of 11 kWh. Under the Added Supply Assumption, an added 170,000 tons per year of graphite (20 percent of China's annual production) is made available, allowing for up to 38.17 kg per vehicle or 80 percent of the requirement of a 60 kWh NMC811 battery. This corresponds to a 48 kWh battery pack. Given performance characteristics associated with smaller battery packs (i.e., reduced range, acceleration, and payload capacity), and consumer aversion to these characteristics (39,40), our results suggest that efforts to meet electrification targets by reducing battery size to this degree may impede EV adoption efforts.

Moreover, we note that our analysis optimistically assumes battery packs scaled for sedan-sized electric vehicles. However, the modern U.S. vehicle market is currently skewed towards heavier SUVs and light trucks (71 percent) versus sedans (29 percent). Accounting for this fleet profile increases aggregate mineral demands given tradeoffs between vehicle range and battery size. At current mineral production rates, accommodating a heavier fleet profile limits EV deployment from 2027-2032 to 4.12 million EVs for the optimal chemistry (NMC 811-only) case and 2.84 million EVs for the market mix case. Incorporating the Added Supply Assumption increases these figures to 18.70 million and 12.90 million EVs respectively, figures that like in the case of considering a sedan-only fleet, still produces an EV sales shortfall in the medium and high sales scenario (which require 28.05 million and 34.62 million EV sales respectively). Consequence, an EV fleet that favors heavier SUVs and pickup trucks will increase the tension between emissions reductions goals envisioned by EV deployment and the limited mineral production available from the US and partner countries.

A second pathway we investigate that realizes the emissions reductions envisioned by the EPA involves HEVs. Could HEVs offer an equivalent emissions benefit envisioned by the EPA proposal? Our model estimates that if policy were to facilitate the exclusive adoption of HEVs rather than NMC 811 EVs, meeting the EPA's emissions reduction goals necessitates at least 189.90 million (93.70 to 219.46 million) HEVs sold between 2027 – 2032. But year-on-year from 2028 onwards (2032 and 2027 onwards in the low and high sales scenarios, respectively), the requisite rate of HEV sales exceeds the total projected light-duty vehicle sales in the US. For example, in the medium sales scenario, we estimate that meeting the EPA's envisioned emissions benefit requires at least 27.72 million HEVs sold in 2030, exceeding the year's estimated light-duty vehicle sales of 15.21 million. Thus, although early HEV sales can achieve rates consistent with realization of the EPA's envisioned emissions benefits, the requisite volume of HEV sales becomes implausibly high in later years.

Nevertheless, we find that HEVs can – under specific sales scenarios – effectively supplement EVs without reducing the potential emissions benefits. In our medium sales scenario, selling at least 2.68 million EVs in 2030 and replacing the remaining year's ICEV sales with HEVs offers a total lifecycle emissions benefit of 60.19 million tons $CO_2e$, which is equivalent to the emissions benefits realized by meeting the EPA's fleet electrification goals using solely EVs. In the high sales scenario, realizing such an emissions benefit requires more stringent thresholds, with at least 2.49 million EVs sold as early as 2027. Across all sales scenarios, our model suggests a minimum of 4.91 million EVs in 2032 are required to enable HEVs to supplement EVs without reducing potential lifecycle emissions benefits. To the extent that increases in mineral production progress at a pace slower than required to realize the EPA's envisioned emissions reduction targets using EVs alone, our results suggest that leveraging a combination of HEVs and EVs may help relax the requisite mineral production required while offering equivalent emissions benefits.

**Limitations and Conclusion**

Given the rapid evolving nature of the EV market, we have endeavored to weight our approach towards an improved value proposition of EVs relative to ICEVs. These assumptions include 1) improvements in battery longevity that obviate the need for battery replacement, 2) grid decarbonization driven by the 2022 Inflation Reduction Act (which improves the emissions proposition of EVs relative to ICEVs and HEVs), 3) battery chemistry market mixes weighted towards cobalt-free LFP batteries, 4) full mineral supply allocation for light-duty EV production absent considering competing use by other electric vehicles (heavy trucks, two-wheelers, off-road utility vehicles) or in other sectors, 5) increased domestic production of limiting critical minerals owing to permitting reform, and 6) mineral requirements that – unlike the current US light-duty vehicle fleet – assume a smaller, lighter vehicle which has lower mineral demands. This approach – we argue – provides reassurance that our findings do not overstate or exaggerate the challenges facing future EV mass adoption. However, we acknowledge opportunities to further build upon our efforts.

Firstly, our model assumes an EV range of 300 miles, a figure that while exceeding the median mileage offered by EVs today, falls short of median mileage offered by current ICEVs (41-43). We note that consideration of EVs that have higher requisite mileage would – all else being equal -- increase the mineral demands associated with widespread electrification, potentially exacerbating the magnitude of disequilibrium enumerated by our model. Increased mineral demands associated with higher range EVs can certainly be tempered by further improvements in EV fuel economy. However, the EPA's emissions proposal prioritizes tailpipe emissions specifically and remains agnostic to EV fuel economy. Put another way, the EPA proposal does not incentivize automakers to improve fuel economy for EVs. Were EV fuel economy to improve over time, current mineral production would – holding range constant – support additional EVs. Moreover, increased EV fuel economy could – through reductions in electricity demand – increase the lifecycle emissions benefits associated with the EPA's policy.

Relatedly, were grid decarbonization to proceed at a more aggressive rate than assumed by our model (44), the lifecycle emissions benefits associated with the EPA proposal would increase. Conversely, the emissions consequences of non-compliance would be more profound. For example, if the carbon intensity of the electrical grid were 90 percent lower (compared to 50 percent assumed by our model) by 2030 relative to 2005, a shortfall in meeting requisite sales targets in the low sales scenario alone would produce the equivalent of 561.39 million tons $CO_2e$ in lost lifecycle emissions benefits. This shortfall is much higher than 284.12 million tons $CO_2e$ estimated by our model which assumes a less aggressive grid decarbonization rate (though one that is within existing estimates).

Secondly, our disequilibrium estimates seek to evaluate the sensitivity of EV deployment to mineral constraints, not accurately simulate real-world trade, market, and policy dynamics. One should not interpret this analysis as suggesting that the US will domestically manufacture all EV battery packs for the domestic market, or that the totality of mineral production across the US and international free trade partners and allies is available for realization of the EPA's U.S. light-duty vehicle electrification goals. In practice, the challenge of procuring sufficient battery mineral supplies is as profound for many US allies as it is for the US and thus poses a key obstacle to global decarbonization efforts (45). Moreover, meeting climate goals necessitates the manufacturing and deployment of other mineral intensive technologies like solar photovoltaic panels, wind turbines, grid-scale energy storage systems, hydrogen electrolyzers, and hydrogen fuel cells—some of which require the same critical minerals as EV batteries. At the same time, investment in mineral production is also accelerating worldwide and tracking every pending or prospective new project in the US and partners abroad is beyond the feasible scope of this study (46).

Third and finally, this analysis only considers availability of upstream mined minerals and does not cover the entirety of the battery cell and battery pack manufacturing chain. Each step of the EV supply chain exhibits its own patterns of geographic industry concentration and faces its own potential bottlenecks. For instance, over half of global refining and processing capacity for mined lithium and cobalt operates in China, as does over three-quarters of lithium-ion battery cell manufacturing (47). Inability to expand these downstream steps of EV battery pack manufacturing may impact EV tax credit eligibility while exposing the EV supply chain to greater volatility, thereby affecting vehicle affordability and mass adoption.

Nevertheless, our results provide compelling evidence that whereas the EV sales targets envisioned by the EPA can deliver significant emissions reductions, constraints in mineral production may impede the extent to which these reductions are realized. Mineral reserves in the US and partners abroad more than suffice to meet the full envelope of EV sales scenarios, but rates of mineral production do not. Specifically, we find that even in the least aggressive sales scenario, existing mineral production supports a maximum of 5.09 million total EVs from 2027 – 2032, a figure that falls well below the minimum 10.21 million EV sales required for compliance with the EPA's proposal. Our model estimates the emissions impact of this shortfall to be 59.54 million tons $CO_2e$. We identify graphite, and to a lesser extent, cobalt as the key limiting minerals for which increased production proportionally expands the number of EV batteries that can be manufactured. We document the precise production thresholds necessary to achieve EV sales complying with the EPA standard. These dynamics and tradeoffs warrant consideration by policymakers as efforts to decarbonize the light-duty vehicle sector accelerate.

**Method**

We assess the viability of the EPA's tailpipe emissions standards in three steps. Further details on our method, underlying assumptions, and leveraged figures are specified in the Supplementary Information section.

*Step 1: Sales volume estimation / scenario construction*

First, we estimate requisite number of EV sales that would satisfy the EPA's standard (11). The standard – which is applicable to new vehicles sold between 2027 and 2032 – necessitates that during, or by the end of this period, the light-duty vehicle fleet achieve an average tailpipe emissions target of 82 grams/mile (g/mile) of $CO_2$ across new vehicle sales. Our model estimates that compliance necessitates EVs constitute no less than 37.82 percent of light-duty- vehicle sales in 2032. Given the proposed standard applies to vehicle sales beginning in 2027, we construct three EV sales volume scenarios (low, medium, and high sales) that each meet the 37.82 percent sales target between 2027 and 2032.

In the low sales scenario, EVs represent 6 percent of annual light-duty vehicle sales (analogous to 2022) until 2032, at which time sales increase to 37.82 percent (48). This represents a lower-bound case wherein vehicle manufacturers do not meaningfully respond – as evidenced by EV sales volume – to the EPA's policy until the year of the compliance deadline itself. In the medium sales scenario, EV sales steadily increase from 6 percent in 2023 to 23.55 percent in 2027 to 37.82 percent in 2032, which is analogous to EVs' market share gradually increasing to satisfy the EPA's adoption targets (see Supplementary Information, Section I for details). In the high sales scenario, EV sales increase to and remain at 37.82 percent between 2027 and 2032, replicating an upper-bound scenario wherein manufacturers engage in proactive rapid compliance within years. Collectively, the low and high sales scenarios encompass the full envelope of possible pathways for vehicle manufacturers to meet current adoption goals, while the medium sales scenario enumerates a more moderate pathway.

*Step 2: Mineral demand / supply estimation*

Having enumerated the requisite number of EV sales required for EPA compliance, we subsequently quantify (in metric tons) the associated mineral demand associated with the batteries required to power these EVs and assess whether these demands can be met using existing supply. We focus on eight minerals used in large quantities in EV batteries, namely Aluminum, Cobalt, Copper, Graphite, Lithium, Manganese, Nickel, and Phosphate.

In scrutinizing mineral demands, we assess – using existing data – the mineral demands of a singular battery chemistry (or a combination thereof, using linear optimization) that would best accommodate the sales volume targets in each sales scenario. We refer to the maximum achievable sales volume using this approach as our result under the optimal chemistry case. We note that these demands vary based on the specific battery chemistry considered (e.g., Nickel Manganese Cobalt (NMC) 811 necessitates more reliance on Nickel and less reliance on Manganese and Cobalt compared to NMC 111) (49). Consequently, we consider the mineral demands associated with six chemistries that overwhelmingly account for the EV battery market. These are NMC 111, NMC 523, NMC 622, NMC 811, NCA, and LFP. Moreover, because the quantity of minerals required also varies – regardless of chemistry – based on vehicle range, we assume equivalent range (i.e., 300 miles) is afforded across all battery chemistries. This range figure, we note, is consistent with longstanding assessments of EVs' viability as a decarbonization pathway and exceeds the current median range of EVs sold today (thereby accommodating potential future improvements in fuel economy (41,50). In addition, we consider a case where the chemical composition of EV batteries sold each year increasingly shifts away from NMC and NCA chemistries towards LFP, with LFP batteries installed in 60 percent of EVs sold in 2030 and thereafter, relative to 36 percent today.

In assessing mineral supplies, we consider two separate categories: production and reserves. Production refers to the amount of a mineral produced from mining on an average annual basis in the United States, US free trade partner countries, and countries party to a mutual defense agreement with the US. Our production estimates additionally include minerals recovered through recycling in the US, but do not include recycled production in overseas markets due to lack of data. Furthermore, we note that our production estimates only consider limited upstream processing such as milling that is performed to convert ore into concentrate forms for ease of transport or sales (i.e., we do not consider midstream or downstream processing such as refining). Reserves, in contrast, refer to the estimated total amount of a mineral geologically occurring within a country that could reasonably be economically extracted. In sum, production refers to short term, annual supply, while reserves refer to long term, total supply. Given local sourcing constraints, our supply estimates also consider minerals that can be sourced from US free trade and/or mutual defense partner countries. These include Australia, Bahrain, Chile, Colombia, Costa Rica, Dominican Republic, El Salvador, Guatemala, Honduras, Israel, Jordan, Mexico, Morocco, Nicaragua, Oman, Panama, Peru, Singapore, S. Korea, Japan, New Zealand, Philippines, Thailand, and all members of the North Atlantic Treaty Organization. In addition to these countries, we also include Austria – a member of the European Union (EU) that, 1) is not included under the other criteria, and 2), has mineral production relevant to EV battery manufacturing. Inclusion of Austria reflects potential realization of an impending minerals-focused free trade agreement between the EU and the United States (51).

Finally, we enumerate how many EV batteries can be manufactured based on a) annual mineral production limits, and b) mineral reserves, and compare these figures to the requisite number of EVs sales necessitated by the EPA proposal under our low, medium, and high scenario. Mismatches between demand and supply in each scenario are quantified annually (i.e., for each year between 2026 and 2033), and in aggregate (2027 through 2032 combined). To ensure EVs are given the maximal advantage, we assume a one-to-one relationship between battery production and EVs (every battery pack is deployed in a vehicle, with none held in inventory or used for repair or replacement) and further assume that mineral supplies are available in their entirety to EV battery production (versus for the manufacture of competing technologies). Furthermore, we presume that no non-battery mineral requirements are constraining for EV deployment.

We additionally consider several sensitivity tests:

- An *Added Supply Assumption* where available production of each mineral increases by an amount equal to 20 percent of the annual production from the top producing country for that respective mineral. In the context of U.S. policies that incentivize 'friend shoring', such an increase could be interpreted in various ways: new production from free trade partners and domestic mine operators, loosened domestic content policies, establishment of free trade agreements with new international partners, boosted secondary production from recycling, or technological advances that increase the productivity of existing mines.

- A *Battery Pack Downsizing Assumption* where EV battery packs are downsized in capacity to hit the desired level of EV deployment in 2032 (5.71 million new EVs sold in the year 2032) under both current mineral production and the Added Supply Assumption.

- A *Heavier Fleet Assumption* in which true EV deployment is skewed towards a mix of 71 percent light trucks and SUVs and 29 percent sedans, as opposed to our default case which considers a fleet of 100 percent sedans. We assume that light trucks and SUVs require a larger battery of approximately 100 kWh to achieve the target range of 300 miles, with correspondingly higher per-pack mineral requirements. We evaluate the potential ceiling to nationwide EV deployment under current and Added Supply mineral constraints for this heavier vehicle fleet.

*Step 3: Emissions impact of disequilibrium*

Here, we determine – as applicable – the emissions impact of being unable to meet each EV sales volume target necessitated by the EPA proposal. To do so, we leverage the GREET model, which is commonly used in vehicle lifecycle emissions analyses (52), to calculate the emissions associated with manufacturing EVs powered by different battery chemistries based on the requisite minerals used for each chemistry. Building on previous literature (53), we subsequently estimate the lifecycle emissions benefit of HEVs and EVs relative to ICEVs year-on-year from 2023 – 2032, accounting for heterogeneity in battery chemistry, rising ICEV and HEV fuel economy, and improvements to the electric grid (see Table 2). Based on the US' target of a 50 percent emissions reduction (relative to 2005) by 2030 (54) – a goal further supported via the enactment of the Inflation Reduction Act (IRA) (55) –, we assume emissions associated with the electric grid decline linearly such that a 50 percent reduction relative to 2005 is achieved in 2030. Regarding HEV fuel economy, we assume an annual improvement rate of 8 percent through 2025 and 10 percent from 2026 – 2032, which is consistent with existing CAFE standards (56). However, owing to diminishing returns on further technical innovation, we impose a capped maximum fuel economy 75 miles per gallon for HEVs[1].

---

[1] PHEVs are excluded from our model given, 1) they offer fuel economy that is – on average – less advantageous than HEVs, 2) are more mineral intensive than HEVs to manufacture, and 3) consistently constitute less than one percent of light duty vehicle sales. We note that this approach is consistent with longstanding mineral supply analysis (57).

**Acknowledgements**

We thank Jessica Dunn, Colin Langan, and Edward Neidermeyer for helpful discussions regarding this work.

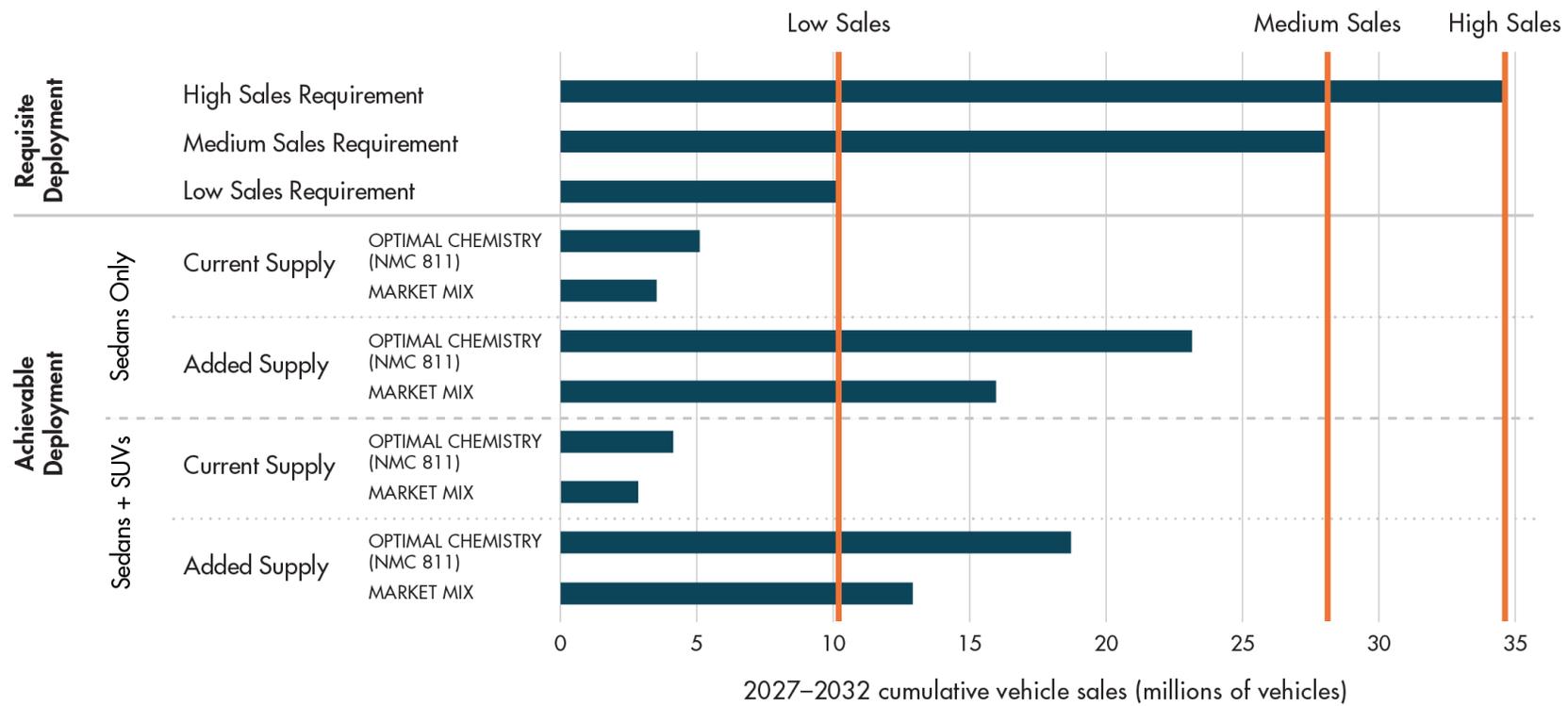

Figure 1: Overview of EV sales scenarios and impact of mineral supply constraints

| | Avg yearly requirement for Biden Admin EV goals (2027–2032) | Max yearly requirement for Biden Admin EV goals (2027–2032) | U.S. avg yearly production, mining only (2018–2022) | U.S. avg yearly production, mining + recyling (2018–2022) | Avg yearly sum of U.S. production, U.S. recycling, and allies and partners' production (2022)** | Potential to increase production through recycling | U.S. avg yearly economy-wide consumption | Recent trend in U.S. economy-wide consumption |
|---|---|---|---|---|---|---|---|---|
| | | | Thousands of metric tons | | | | | |
| Aluminum | Low: 64 Middle: 176 High: 218 | 221 | 24* | 1 524 | 26 524 | High | 4 600 | Steady → |
| Cobalt | Low: 11 Middle: 29 High: 36 | 37 | 0.6 | 2.8 | 22 | High | 9 | Steady → |
| Copper | Low: 43 Middle: 118 High: 145 | 147 | 1 240 | 1 399 | 11 289 | High | 1 832 | Increasing ↑ |
| Graphite | Low: 96 Middle: 264 High: 326 | 331 | Minimal | Minimal | 48 | Low | 50 | Increasing ↑ |
| Lithium | Low: 11 Middle: 29 High: 36 | 37 | 1 | 1 | 102 | Low | 2 | Will increase sharply ↑↑ |
| Manganese | Low: 11 Middle: 29 High: 36 | 37 | Minimal | Minimal | 3 530 | Low | 754 | Steady → |
| Nickel | Low: 83 Middle: 229 High: 283 | 287 | 17 | 123 | 933 | High | 220 | Increasing ↑ |
| Phosphate | Low: 0 Middle: 0 High: 0 | 0 | 23 000 | 23 000 | 84 950 | Low | 25 000 | Steady → |

*U.S. mine production from 2021 only     **Considers countries with a Free Trade Agreement or mutual defense treaty with the United States

■ Sufficient for all cases   ■ Sufficient for Middle Case   ■ Sufficient for Low Case   ■ Insufficient in all cases
■ Processing, refining, or production supply chains highly concentrated in China

Figure 2a: Overview of mineral demands versus available supply (Optimal chemistry – NMC 811)

| | Avg yearly requirement for Biden Admin EV goals (2027–2032) | Max yearly requirement for Biden Admin EV goals (2027–2032) | U.S. avg yearly production, mining only (2018–2022) | U.S. avg yearly production, mining + recyling (2018–2022) | Avg yearly sum of U.S. production, U.S. recycling, and allies and partners' production (2022)** | Potential to increase production through recycling | U.S. avg yearly economy-wide consumption | Recent trend in U.S. economy-wide consumption |
|---|---|---|---|---|---|---|---|---|
| | Thousands of metric tons | | | | | | | |
| Aluminum | Low: 94 Middle: 257 High: 316 | 319 | 24* | 1 524 | 26 524 | High | 4 600 | Steady → |
| Cobalt | Low: 6 Middle: 16 High: 20 | 24 | 0.6 | 2.8 | 22 | High | 9 | Steady → |
| Copper | Low: 55 Middle: 152 High: 187 | 188 | 1 240 | 1 399 | 11 289 | High | 1 832 | Increasing ↑ |
| Graphite | Low: 140 Middle: 384 High: 473 | 477 | Minimal | Minimal | 48 | Low | 50 | Increasing ↑ |
| Lithium | Low: 14 Middle: 39 High: 49 | 49 | 1 | 1 | 102 | Low | 2 | Will increase sharply ↑↑ |
| Manganese | Low: 5 Middle: 14 High: 17 | 20 | Minimal | Minimal | 3 530 | Low | 754 | Steady → |
| Nickel | Low: 36 Middle: 101 High: 127 | 147 | 17 | 123 | 933 | High | 220 | Increasing ↑ |
| Phosphate | Low: 36 Middle: 98 High: 120 | 126 | 23 000 | 23 000 | 84 950 | Low | 25 000 | Steady → |

*U.S. mine production from 2021 only    **Considers countries with a Free Trade Agreement or mutual defense treaty with the United States

■ Sufficient for all cases   ■ Sufficient for Middle Case   ■ Sufficient for Low Case   ■ Insufficient in all cases
■ Processing, refining, or production supply chains highly concentrated in China

Figure 2b: Overview of mineral demands versus available supply (Market mix)

|  |  | 2027 | 2028 | 2029 | 2030 | 2031 | 2032 |
|---|---|---|---|---|---|---|---|
| All sales scenarios | Projected light-duty vehicle sales | 15,478,700 | 15,330,200 | 15,268,900 | 15,210,400 | 15,144,000 | 15,102,000 |
| Low sales scenario | # of EVs desired | 911,257 | 902,515 | 898,906 | 895,462 | 891,553 | 5,711,810 |
|  | # of EVs possible (Production) | 848,804 | 848,804 | 848,804 | 848,804 | 848,804 | 848,804 |
|  | Optimal EV battery chemistry | NMC 811 | NMC 811 | NMC 811 | NMC 811 | NMC 811 | NMC 811 |
|  | Emissions shortfall from lack of EVs (tons $CO_2e$) | 838,865 | 701,400 | 634,080 | 574,939 | 509,334 | 56,281,611 |
| Medium sales scenario | # of EVs desired | 3,645,234 | 4,047,671 | 4,467,348 | 4,884,425 | 5,295,399 | 5,711,810 |
|  | # of EVs possible (Production) | 848,804 | 848,804 | 848,804 | 848,804 | 848,804 | 848,804 |
|  | Optimal EV battery chemistry | NMC 811 | NMC 811 | NMC 811 | NMC 811 | NMC 811 | NMC 811 |
|  | Emissions shortfall from lack of EVs (tons $CO_2e$) | 37,561,279 | 41,773,389 | 45,795,416 | 49,728,490 | 52,979,047 | 56,281,611 |
| High sales scenario | # of EVs desired | 5,854,284 | 5,798,119 | 5,774,935 | 5,752,809 | 5,727,696 | 5,711,810 |
|  | # of EVs possible (Production) | 848,804 | 848,804 | 848,804 | 848,804 | 848,804 | 848,804 |
|  | Optimal EV battery chemistry | NMC 811 | NMC 811 | NMC 811 | NMC 811 | NMC 811 | NMC 811 |
|  | Emissions shortfall from lack of EVs (tons $CO_2e$) | 67,232,948 | 64,632,153 | 62,343,920 | 60,429,055 | 58,129,658 | 56,281,611 |

Table 1: Model results for EV sales scenarios

| | 2027 | 2028 | 2029 | 2030 | 2031 | 2032 |
|---|---|---|---|---|---|---|
| ICEV Fuel Economy | 60.00 | 61.20 | 62.50 | 63.70 | 65.10 | 66.40 |
| HEV Fuel Economy | 73.39 | 75 | 75 | 75 | 75 | 75 |
| EV Fuel Economy | 114 | 114 | 114 | 114 | 114 | 114 |
| Electric Grid Emissions Rate (g $CO_2$e/kWh) | 309.78 | 304.12 | 298.56 | 293.10 | 287.74 | 282.48 |
| Lifecycle emissions – ICEV (tons $CO_2$e/vehicle) | 40.21 | 39.58 | 38.92 | 38.34 | 37.69 | 37.11 |
| Lifecycle emissions – HEV (tons $CO_2$e/vehicle) | 36.73 | 36.16 | 36.16 | 36.16 | 36.16 | 36.16 |
| Lifecycle emissions – EV NMC 111 (tons $CO_2$e/vehicle) | 26.34 | 26.08 | 25.83 | 25.58 | 25.34 | 25.09 |
| Lifecycle emissions – EV NMC 523 (tons $CO_2$e/vehicle) | 27.12 | 26.86 | 26.61 | 26.36 | 26.11 | 25.87 |
| Lifecycle emissions – EV NMC 622 (tons $CO_2$e/vehicle) | 27.22 | 26.97 | 26.71 | 26.46 | 26.22 | 25.98 |
| Lifecycle emissions – EV NMC 811 (tons $CO_2$e/vehicle) | 26.77 | 26.52 | 26.26 | 26.01 | 25.77 | 25.53 |
| Lifecycle emissions – EV NCA (tons $CO_2$e/vehicle) | 27.10 | 26.85 | 26.59 | 26.34 | 26.10 | 25.86 |
| Lifecycle emissions – EV LFP (tons $CO_2$e/vehicle) | 26.06 | 25.81 | 25.55 | 25.30 | 25.06 | 24.82 |
| Lifecycle emissions – EV weighted average (tons $CO_2$e/vehicle) | 26.54 | 26.26 | 25.95 | 25.75 | 25.52 | 25.28 |

Table 2: Lifecycle emissions by powertrain and chemistry

**Supplementary Information**

Here, we provide an overview of our model, assumptions made, and outcomes enumerated. In Section I, we estimate the requisite EV market share required to comply with the EPA's tailpipe emissions standard. In Section II, we assess available mineral supply available domestically and among US allies. In Section III, we consider EV battery manufacturing capacity based on available mineral supply and assess the magnitude (if any) of disequilibrium between mineral supply and demand. In Section IV, we apply sensitivity tests to assess the robustness of our model's assumptions. In Section V, we scrutinize the emissions impact of disequilibrium between mineral supply and demand. Finally, in Section VI, we assess the viability of pathways that reduce disequilibrium in critical mineral.

**SECTION I: Electric Vehicle Sales Projections**

We begin by estimating the requisite market penetration of electric vehicles (EVs) to realize the EPA's industry-wide average tailpipe emissions target for the light-duty fleet of 82 grams/mile (g/mile) of $CO_2$ by 2032 (1). To do so, we leverage existing data on emissions associated with fuel usage for internal combustion engine vehicles (ICEVs) and hybrid electric vehicles (HEVs). In line with previous efforts (2), we assume an average emissions rate of 73 grams $CO_2$ per megajoule and fuel economies of 66.40 and 75 miles per gallon for ICEVs and HEVs respectively. These numbers are summarized in Table 1.1. For the purposes of our analysis, we presume that the administration's classification of EVs exclusively denotes battery electric vehicles, which have zero emissions associated with fuel usage owing to their reliance on electricity. Consequently, we find that EVs must achieve 37.82 percent market penetration to realize a light-duty fleet-wide average of 82 g/mile $CO_2$. Leveraging this 37.82 percent target, we project electric vehicle sales out to 2032 in a two-step process.

|  | Average Emissions Rate (grams $CO_2$ per megajoule) | Fuel Economy (miles per gallon) |
|---|---|---|
| ICEVs | 73 | 66.40 |
| HEVs | | 75 |

Table 1.1: Average emissions rate and fuel economy of ICEVs and HEVs

First, we project total light-vehicle sales (including vehicles with non-electric powertrains) through a linear projection of data from the U.S. Energy Information Administration (3) and the National Automobile Dealers Association (4). This is in line with historical data from the Bureau of Economic Analysis (5), which demonstrates the cyclical nature of new car sales.

Next, electric vehicle sales are projected as a percentage of total new car sales. Here, we create three different scenarios that serve as the low, medium and high sales scenarios respectively. In order to develop these three scenarios, we first examine historical electric vehicle sales as a percentage of total light-vehicle sales by leveraging data from the U.S. Department of Energy (6). Most notably, electric vehicle sales were 5.89 percent of all new car sales in 2022. Our low, medium, and high sales scenarios are described as follows:

Low Sales Scenario: For the low sales scenario, we maintain the 2022 proportion of electric vehicles (i.e., 5.89 percent of total new car sales) until 2032, upon which we increase the value to 37.82 percent to meet the Biden administration's 2032 target.

Medium Sales Scenario: To achieve the 37.82 percent target by 2032, electric vehicle sales as a percentage of total new light-vehicle sales need to double approximately three times from the 5.89 percent in 2022. We space out these three doublings for 2024, 2027, and 2032 with electric vehicle sales percentages of 11.77, 23.55, and 37.82 percent respectively. We model steady increases in between each doubling. For example, there is an 11.77 percent difference between electric vehicle sales percentages for 2024 and 2027. This 11.77 percent increase is evenly spaced out such that each year between 2024 and 2027 sees an incremental 3.92 percent increase compared to the prior year (11.77 percent / 3).

High Sales Scenarios: Our high sales scenario models out the optimistic scenario that 37.82 percent of new car sales would be electric vehicles starting immediately from 2023. This 37.82 percent value is maintained out to 2032.

These choices of Low and High sales scenarios should not be interpreted as realistic or plausible future projections. Rather, these scenarios capture the entire possible envelope of pathways by which U.S. electric vehicle sales could attain the 2032 Biden Administration targets. The respective electric vehicle sales percentages for the low, medium and high sales scenarios are then multiplied by the projected total new car sales to yield our estimates for electric vehicle sales out to 2032. Numbers for these projections are summarized in Table 1.2.

| | 2022 | 2023 | 2024 | 2025 | 2026 | 2027 | 2028 | 2029 | 2030 | 2031 | 2032 |
|---|---|---|---|---|---|---|---|---|---|---|---|
| Total New Car Sales | 13,754,300 | 14,600,000 | 15,492,600 | 15,456,400 | 15,543,100 | 15,478,700 | 15,330,200 | 15,268,900 | 15,210,400 | 15,144,000 | 15,102,000 |
| Low Sales Scenario (EV Sales %) | 5.89% | 5.89% | 5.89% | 5.89% | 5.89% | 5.89% | 5.89% | 5.89% | 5.89% | 5.89% | 37.82% |
| Low Sales Scenario (EV Sales) | 809,739 | 859,527 | 912,076 | 909,945 | 915,049 | 911,257 | 902,515 | 898,906 | 895,462 | 891,553 | 5,711,810 |
| Medium Sales Scenario (EV Sales %) | 5.89% | 8.83% | 11.77% | 15.70% | 19.62% | 23.55% | 26.40% | 29.26% | 32.11% | 34.97% | 37.82% |
| Medium Sales Scenario (EV Sales) | 809,739 | 1,289,290 | 1,824,151 | 2,426,519 | 3,050,162 | 3,645,029 | 4,047,671 | 4,467,348 | 4,884,425 | 5,295,399 | 5,711,810 |
| High Sales Scenario (EV Sales %) | 5.89% | 37.82% | 37.82% | 37.82% | 37.82% | 37.82% | 37.82% | 37.82% | 37.82% | 37.82% | 37.82% |
| High Sales Scenario (EV Sales) | 809,739 | 5,521,946 | 5,859,541 | 5,845,850 | 5,878,641 | 5,854,284 | 5,798,119 | 5,774,935 | 5,752,809 | 5,727,696 | 5,711,810 |

Table 1.2:  EV sales projections (low, medium and high sales scenarios)

**SECTION II: Mineral Supply Projections**

Next, we estimate the supply of minerals under current mineral production and estimates of allies' reserves.

Note that U.S. partners / allies are defined as countries with which the U.S. either has free trade agreements or treaties in place. We did not consider the Rio Treaty due to its limited relevance. These countries are listed below:

- <u>U.S. Free Trade Agreements</u>: Australia, Bahrain, Canada, Chile, Colombia, Costa Rica, Dominican Republic, El Salvador, Guatemala, Honduras, Israel, Jordan, S. Korea, Mexico, Morocco, Nicaragua, Oman, Panama, Peru, Singapore
- <u>U.S. Treaty Allies</u>: NATO, S. Korea, Japan, Australia, New Zealand, Philippines, Thailand

We further include Austria - a member of the European Union (EU) that (1) is not included under the other criteria, and 2), has mineral production relevant to EV battery manufacturing. Inclusion of Austria reflects potential realization of an impending minerals-focused free trade agreement between the EU and the United States.

Commencing mineral supply projections entails first defining metrics used for current production rates and estimated reserves. Production rates measure the current realized ability to produce a flow of minerals either from mine or recycling production, whereas reserves refer to the estimated total stock of geologic mineral deposits currently assessed as economic and reflect today's overall capacity / potential for cumulative mineral production irrespective of time. To increase production, recycling capacities can be enlarged, existing mines can be expanded, or new mines can be developed to access existing reserves; to increase reserves, mineral exploration can be conducted to discover additional geologic deposits, or improved market conditions and / or technological innovations can upgrade previously uneconomic deposits into economically viable reserves. Consequently, the reserve estimates we leverage should be considered conservative, lower bounds given the potential increases due to exploration activities likely to occur over the study's period.

***Part A. Mineral Production Estimates***

Mineral production rates for the U.S. are calculated as averages of annual mine and recycling production data from 2018 – 2022 unless otherwise noted, while foreign production rates are taken directly from mine production data from 2022. In both cases, production data from 2022 are estimates. We provide production rates from both mining and recycling for the U.S. when applicable, while production rates for U.S. allies / partners only include mine production due to the lack of availability of data on recycling-based production. Mineral production rates for both the U.S. and allies / partners are summarized below in Table 2.1. Note that the annual production requirements in the leftmost column are taken from mineral demand projections in Section II. The annual requirement for Biden's 2032 goal is calculated assuming that only EVs with the NMC 811 battery chemistry are produced, which – our model estimates - is the optimal battery chemistry to maximize production given current mineral supplies (see Section III for details).

| | Annual Requirement for Biden's 2032 Goal | U.S. Annual Production (Mining) | U.S. Annual Production (Mining + Recycling) | U.S. Allies + Partners Annual Production (Mining) |
|---|---|---|---|---|
| Lithium | Low: 11,000 Medium: 29,000 High: 36,000 | 1,000[a] | 1,000[b] | 101,000[c] |
| Cobalt | Low: 11,000 Medium: 29,000 High: 36,000 | 600[d] | 2,800[e] | 19,000[f] |
| Nickel | Low: 83,000 Medium: 229,000 High: 283,000 | 17,000[g] | 123,000[h] | 810,000[i] |
| Manganese | Low: 11,000 Medium: 29,000 High: 36,000 | None[j] | None[j] | 3,530,000[k] |
| Graphite | Low: 96,000 Medium: 264,000 High: 326,000 | None[l] | None[m] | 48,000[n] |
| Aluminum | Low: 64,000 Medium: 176,000 High: 218,000 | 24,000[o] | 1,524,000[p] | 25,000,000[q] |
| Copper | Low: 43,000<br>Medium: 118,000<br>High: 145,000 | 1,240,000[r] | 1,399,000[s] | 9,890,000[t] |
| Phosphate | Low: 0<br>Medium: 0<br>High: 0 | 23,000,000[u] | 23,000,000[v] | 61,950,000[w] |

Table 2.1: Mineral production estimates (metric tons)

[a] The only active U.S. lithium mine is Silver Peak in Nevada, which produces just under 1,000 metric tons (mt) on an annual basis (7).

[b] There is no lithium production from recycling (8), so our total figure for production from both mining and recycling sources is the same as that of mining – or 1,000 mt.

[c] Our estimates for U.S. allies and partners included 2022 mining values for Chile, Australia, Canada, and Portugal respectively: 39,000 mt + 61,000 mt + 500 mt + 600 mt = 101,100 mt (8). This value was rounded down to 101,000 mt.

[d] Annual cobalt mine production for the U.S. was calculated as a 5-year average for 2018-2022. Annual values for these years were 480 mt, 500 mt, 600 mt, 650 mt, and 800 mt respectively (8). This yielded an average estimate of 606 mt. We then rounded this value down to 600 mt.

[e] Annual cobalt recycling production for the U.S. was calculated as a 5-year average for 2018-2022. Annual values for these years were 2,750 mt, 2,750 mt, 2,010 mt, 1,800 mt, and 1,900 mt (8). This yielded an average recycling production estimate of 2,242 mt. We rounded this down to 2,200 mt. Adding this to our mining estimate, we arrived at a total estimate of 600 mt + 2,200 mt = 2,800 mt.

[f] Our estimates for U.S. allies and partners included 2022 mining values for Australia, the Philippines, Canada, Turkey, and Morocco respectively: 5,900 mt + 3,800 mt + 3,900 mt + 2,700 mt + 2,300 mt = 18,600 mt (8). This value was rounded up to 19,000 mt.

[g] Annual nickel mine production for the U.S. was calculated as a 5-year average for 2018-2022. Annual values for these years were 17,600 mt, 13,500 mt, 16,700 mt, 18,400 mt, and 18,000 mt respectively (8). This yielded an average estimate of 16,840 mt. We then rounded this value up to 17,000 mt.

[h] Annual nickel recycling production for the U.S. was calculated as a 5-year average for 2018-2022. Annual values for these years were 123,000 mt, 111,000 mt, 100,000 mt, 100,000 mt, and 96,000 mt respectively (8). This yielded an average recycling production estimate of 106,000 mt. Adding this to our mining estimate, we arrived at a total estimate of 17,000 mt + 106,000 mt = 123,000 mt.

[i] Our estimates for U.S. allies and partners included 2022 mining values for Australia, Canada, New Caledonia (France), and the Philippines respectively: 160,000 mt + 130,000 mt + 190,000 mt + 330,000 mt = 810,000 mt (8).

[j] Manganese has not been mined domestically since the 1970s partly due to low ore grades of less than 20 percent compared with average metallurgical grades of 44 percent, so we estimated a value of 0 mt for U.S. production from both mining and recycling (8).

[k] Our estimates for U.S. allies and partners included 2022 mining values for Australia and Mexico respectively: 3,300,000 mt + 230,000 mt = 3,530,000 mt (8).

[l] Graphite is produced both naturally and synthetically. However, most synthetic graphite production is consumed in the steelmaking industry and is not of battery-grade quality. Natural graphite has not been mined domestically since the 1950s, so we estimate a value of 0 mt of U.S. production from mining.

[m] There is no domestic production of natural graphite from recycling. Therefore, our total figure for production from both mining and recycling sources is the same as that of mining – or 0 mt.

[n] Our estimates for U.S. allies and partners included 2022 mining values for natural graphite from Austria, Canada, Germany, South Korea, Mexico, Norway, and Turkey respectively: 500 mt + 15,000 mt + 250 mt + 17,000 mt + 1,900 mt + 10,000 mt + 2,900 mt = 47,550 mt (8). This value was rounded up to 48,000 mt.

[o] Aluminum mine production was estimated with respect to bauxite mine production, which is the major aluminum-bearing ore and a precursor for aluminum production. Due to restrictions on proprietary data, mine production rates could not be calculated as averages of multiple years and instead directly used data from 2021 bauxite mine production which was 96,000 mt (10). Equivalent aluminum production was then derived by leveraging the following conversion equation: *Aluminum Equivalent = Bauxite ÷ 4*. This makes the simplifying approximation that 4 tons of dried bauxite are required to produce 2 tons of alumina, which in turn can be used to produce 1 ton of aluminum – yielding a 4:1 conversion factor for bauxite to aluminum. Therefore, we estimated annual U.S. mine production to be 96,000 mt ÷ 4 = 24,000 mt.

[p] Annual aluminum recycling production values considered direct recycling of aluminum (as opposed to bauxite or alumina) which for 2018, 2019, 2020, 2021, and 2022 were 1,570,000 mt, 1,540,000 mt, 1,420,000 mt, 1,520,000 mt, and 1,500,000 mt, respectively. This yielded a 5-year average across 2018-2022 of 1,510,000 mt. We rounded this down to 1,500,000 mt of recycled aluminum production, resulting in a total of 24,000 mt + 1,500,000 mt = 1,524,000 mt of aluminum production from both mining and recycling.

[q] Annual aluminum mine production across U.S. allies and partners was also calculated with respect to bauxite mine production and was derived from 2022 mining values for Australia: 100,000,000 mt of bauxite (8). We leveraged the 4:1 bauxite to aluminum conversion factor to arrive at 25,000,000 mt.

[r] Annual copper mine production for the U.S. was calculated as a 5-year average for 2018-2022. Annual values for these years were 1,220,000 mt, 1,260,000 mt, 1,200,000 mt, 1,230,000 mt, and 1,300,000 mt respectively (8). This yielded an average estimate of 1,242,000 mt which we then round down to 1,240,000 mt.

[s] Annual copper recycling production for the U.S. was calculated as a 5-year average for 2018-2022. Annual values for these years were 141,000 mt, 166,000 mt, 160,000 mt, 170,000 mt, and 160,000 mt respectively (8). This yielded an average secondary production estimate of 159,400 mt. We then rounded this down to 159,000 mt. Adding this to our mining estimate, we arrived at a total estimate of 1,240,000 mt + 159,000 mt = 1,399,000 mt.

[t] Our estimates for U.S. allies and partners included 2022 mining values for Australia, Canada, Chile, Mexico, Peru, and Poland respectively: 830,000 mt + 530,000 mt + 5,200,000 mt + 740,000 mt + 2,200,000 mt + 390,000 mt = 9,890,000 mt (8).

[u] Annual phosphate mine production for the U.S. was calculated as a 5-year average for 2018-2022. Annual values for these years were 25,800,000 mt, 23,300,000 mt, 23,500,000 mt, 21,600,000 mt, and 21,000,000 mt respectively (8). This yielded an average estimate of 23,040,000 mt. We then rounded this value down to 23,000,000 mt.

[v] There is no domestic phosphate production from recycling, so our total estimate remains 23,000,000 mt.

[w] Our estimates for U.S. allies and partners included 2022 mining values for Australia, Finland, Israel, Jordan, Mexico, Morocco, Peru, and Turkey respectively: 2,500,000 mt + 1,000,000 mt + 3,000,000 mt + 10,000,000 mt + 450,000 mt + 40,000,000 mt + 4,200,000 mt + 800,000 mt = 61,950,000 mt (8).

***Part B. Mineral Reserve Estimates***

Reserve figures for both the U.S. and its allies / partners are calculated with respect to 2023 estimates and are shown below in Table 2.2. Reserve figures for Australia are with respect to Joint Ore Reserves Committee-compliant or equivalent reserves. Note that the cumulative 2027-2032 requirements are taken from mineral demand projections in Section II.

|  | Cumulative 2027-2032 Needs for Biden's 2032 Goal | U.S. Estimated Reserves | U.S. Allies + Partners Estimated Reserves |
|---|---|---|---|
| Lithium | Low: 64,000<br>Medium: 176,000<br>High: 218,000 | 1,000,000[x] | 14,090,000[y] |
| Cobalt | Low: 64,000<br>Medium: 176,000<br>High: 218,000 | 69,000[z] | 1,199,000[aa] |
| Nickel | Low: 500,000<br>Medium: 1,370,000<br>High: 1,700,000 | 370,000[ab] | 23,600,000[ac] |
| Manganese | Low: 64,000<br>Medium: 176,000<br>High: 218,000 | Negligible[ad] | 140,000,000[ae] |
| Graphite | Low: 577,000<br>Medium: 1,590,000<br>High: 1,960,000 | Negligible[af] | 95,500,000[ag] |
| Aluminum | Low: 385,000<br>Medium: 1,060,000<br>High: 1,310,000 | 5,000,000[ah] | 425,000,000[ai] |
| Copper | Low: 257,000<br>Medium: 705,000<br>High: 870,000 | 44,000,000[aj] | 384,600,000[ak] |
| Phosphate | Low: 0<br>Medium: 0<br>High: 0 | 1,000,000,000[al] | 52,474,000,000[am] |

Table 2.2: Mineral reserve estimates (metric tons)

[x] The U.S. is estimated to have lithium reserves of 1,000,000 mt (8).

[y] Our estimates for U.S. allies and partners included reserve values for Chile, Australia, Canada, and Portugal respectively: 9,300,000 mt + 3,800,000 mt + 930,000 mt + 60,000 mt = 14,090,000 mt (8).

[z] The U.S. is estimated to have cobalt reserves of 69,000 mt (8).

[aa] Our estimates for U.S. allies and partners included reserve values for Australia, the Philippines, Canada, Turkey, and Morocco respectively: 670,000 mt + 260,000 mt + 220,000 mt + 36,000 mt + 13,000 mt = 1,199,000 mt (8).

[ab] The U.S. is estimated to have nickel reserves of 370,000 mt (8).

[ac] Our estimates for U.S. allies and partners included reserve values for Australia, Canada, New Caledonia (France), and the Philippines respectively: 9,500,000 mt + 2,200,000 mt + 7,100,000 mt + 4,800,000 mt = 23,600,000 mt (8).

[ad] The U.S. is estimated to have a negligible amount of manganese reserves partly due to low ore grades of less than 20 percent compared with average metallurgical grades of 44 percent (8).

[ae] Our estimates for U.S. allies and partners included reserve values for Australia and Mexico respectively: 135,000,000 mt + 5,000,000 mt = 140,000,000 mt (8).

[af] The U.S. does not currently report graphite reserves significant enough to be specified as of the most recent report by the U.S. Geological Survey and are taken to be negligible (8). However, some commercial exploratory efforts are considering domestic natural graphite mine development so reserves figures are likely to be revised upward soon.

[ag] Our estimates for U.S. allies and partners included reserve values for South Korea, Mexico, Norway, and Turkey respectively: 1,800,000 mt + 3,100,000 mt + 600,000 mt + 90,000,000 mt = 95,500,000 mt (8). While Austria, Canada, and Germany, report graphite production, their corresponding reserves are small enough to be unspecified as of the most recent report by the U.S. Geological Survey and are then taken to be negligible.

[ah] Aluminum reserves were estimated with respect to bauxite reserves, which is the major aluminum-bearing ore and a precursor for aluminum production. The U.S. is estimated to have bauxite reserves of 20,000,000 mt (8). Equivalent aluminum reserves were then derived by leveraging the following conversion equation: *Aluminum Equivalent = Bauxite ÷ 4*. This makes the simplifying approximation that 4 tons of dried bauxite are required to produce 2 tons of alumina, which in turn can be used to produce 1 ton of aluminum – yielding a 4:1 conversion factor for bauxite to aluminum. Therefore, we estimated U.S. aluminum reserves to be 20,000,000 mt ÷ 4 = 5,000,000 mt.

**SECTION III: Mineral Demand Estimation**

In this section, we execute the following steps:

- First, we outline the mineral content for different battery chemistries, standardized for a light-duty four-door sedan with a 300-mile range.

- Based on mineral production limits, we consider the number of EV batteries that can be produced in the "optimal chemistry" scenario. In this scenario, we identify the battery chemistry that produces the greatest number of EV batteries and assume only EV batteries of this chemistry are produced.

- We also consider a hypothetical "market mix" scenario. This scenario assumes a more realistic assumption that the US EV fleet expands following an evolving mix of several battery chemistries that shift increasingly towards LFP battery packs.

*Part A. Mineral Content for Different Battery Chemistries*

Here, we assess the mineral content for different battery chemistries. To begin, we first identify the mineral content that is required for the following six battery chemistries: NMC111, NMC532, NMC622, NMC811, NCA, and LFP. For the purposes of our analysis, we standardize for a light-duty four-door sedan with a 300-mile range across each battery chemistry using data on recent EV models' battery weights and ranges (11). We note that although consideration of NMC 111 entails incorporation of an upper, middle and lower bound mineral intensity, this chemistry type accounts for a minor share of the electric vehicle battery market between the present day and 2032. Consequently, this heterogeneity does not significantly affect overall mineral demand calculations.

Mineral contents for the six different battery chemistries are summarized in Table 3.1 below. Note that the values shown for NMC111 are averages of the lower and upper bound mineral intensities.

|  | NMC 111 | NMC 523 | NMC 622 | NMC 811 | NCA | LFP |
|---|---|---|---|---|---|---|
| Lithium | 9.90 (9.20 − 10.61) | 8.94 | 8.47 | 6.28 | 8.35 | 8.78 |
| Cobalt | 26.58 (23.00 − 30.16) | 14.06 | 15.53 | 6.28 | 2.78 | 0.00 |
| Nickel | 26.13 (22.23 − 30.02) | 35.78 | 45.17 | 49.01 | 59.87 | 0.00 |
| Manganese | 24.79 (21.47 − 28.11) | 20.45 | 14.12 | 6.28 | 0.00 | 0.00 |
| Graphite | 60.06 | 67.72 | 70.58 | 56.55 | 61.26 | 96.54 |
| Aluminum | 44.72 | 44.72 | 46.59 | 37.70 | 41.77 | 64.36 |
| Copper | 25.56 | 25.56 | 26.82 | 25.13 | 23.67 | 38.03 |
| Phosphate | 0.00 | 0.00 | 0.00 | 0.00 | 0.00 | 36.57 |

Table 3.1: Mineral content for battery chemistries for a standard sedan (kg)

***Part B. Estimates of Possible Number of EV Batteries (Optimal Chemistry Scenario)***

To determine the optimal battery mix and the number of EV batteries that can be produced, we estimate the number of EV batteries that can be manufactured given annual production limits and mineral reserves for each battery chemistry. The total number of batteries that can be manufactured given each mineral $i$ and each battery chemistry $j$ was calculated as follows:

$$EV\ Batteries_{\ i} = \frac{U.S.\ Production + U.S.\ Recycling + Allies\ Production}{Mineral\ Content_{\ i,j}}$$

Here, mineral $i$ refers to eight battery minerals (lithium, cobalt, nickel, manganese, graphite, aluminum, copper, and phosphate), and battery chemistry $j$ encompasses six different battery chemistries (NMC111, NMC523, NMC622, NMC811, NCA, and LFP). The figures are shown in Table 3.2 below. Bolded numbers indicate that the mineral is the 'limiting factor', determining the maximum number of EV batteries that can be produced.

|  | NMC111 | NMC523 | NMC622 | NMC811 | NCA | LFP |
|---|---|---|---|---|---|---|
| Lithium | 10,299,719 | 11,403,261 | 12,042,475 | 16,233,379 | 12,209,749 | 11,622,660 |
| Cobalt | 820,201 | 1,550,925 | 1,403,882 | 3,469,487 | 7,828,604 | n/a |
| Nickel | 35,703,867 | 26,076,574 | 20,653,729 | 19,036,881 | 15,583,715 | n/a |
| Manganese | 142,396,690 | 172,655,987 | 250,058,441 | 561,802,223 | n/a | n/a |
| Graphite | **799,227** | **708,749** | **680,046** | **848,804** | **783,513** | **497,226** |
| Aluminum | 593,058,992 | 593,058,992 | 569,366,478 | 703,552,510 | 635,002,717 | 412,138,284 |
| Copper | 441,725,425 | 441,725,425 | 420,890,077 | 449,163,265 | 476,940,685 | 296,851,153 |
| Phosphate | n/a | n/a | n/a | n/a | n/a | 2,323,164,645 |

Table 3.2: Number of EV batteries given each battery chemistry

To maximize the production of EV batteries given the total mineral supply, the optimal battery chemistry is NMC811, producing 848,804 batteries.

From there, we could compare the maximum number of EV batteries that can be manufactured to the requisite number of EVs sales necessitated by the EPA proposal under our low, medium and high sales scenarios.  This allows us to calculate the shortfall of EV batteries as follows:

$$EV\ Batteries\ Shortfall_{k,t} = Desired\ EVs_{k,t} - Possible\ EVs_{k,t}$$

Here, EV sales scenario $k$ refers to the low, medium, or high sales scenario; and time $t$ ranges from 2023 to 2032.

***Part C. Battery Mineral Demand Projections (Hypothetical Market Share Scenario)***

Here, we project demand based on a hypothetical mix of battery chemistries. In this case, we consider the relative market shares for the six different battery chemistries. Projections are derived from a scenario in which LFP batteries are expected to derive a significantly larger share of the market (12). This represents an optimistic case in which LFP batteries possess a high potential to replace lithium-ion battery chemistries. This conservatively biases our demand estimates lower for cobalt, nickel, and manganese, as LFP batteries require less kilograms per battery when compared to other battery chemistries. Full market share projections for all six battery chemistries out to 2032 are shown below in Table 3.3.

|  | 2024 | 2025 | 2026 | 2027 | 2028 | 2029 | 2030 | 2031 | 2032 |
|---|---|---|---|---|---|---|---|---|---|
| NMC111 | 2.6% | 2.2% | 1.9% | 1.6% | 1.3% | 1.0% | 0.8% | 0.8% | 0.8% |
| NMC523 | 4.9% | 3.3% | 2.9% | 2.5% | 2.2% | 1.8% | 1.5% | 1.5% | 1.5% |
| NMC622 | 14.2% | 14.3% | 13.8% | 13.2% | 12.7% | 12.1% | 11.3% | 11.3% | 11.3% |
| NMC811 | 8.6% | 9.8% | 10.0% | 10.2% | 10.3% | 10.4% | 10.3% | 10.2% | 10.2% |
| NCA | 27.4% | 25.2% | 23.2% | 21.3% | 19.4% | 17.7% | 15.8% | 15.8% | 15.8% |
| LFP | 42.3% | 45.3% | 48.2% | 51.2% | 54.1% | 57.1% | 60.0% | 60.0% | 60.0% |

Table 3.3: Market share for battery chemistries (%)

On an individual basis, the mineral demand for mineral $i$, battery chemistry $j$, and EV sales scenario $k$ at time $t$ can be calculated as:

$$Mineral\ Demand_{i,j,k,t} = Mineral\ Content_{i,j} \times Market\ Share_{j,t} \times EV\ Sales_{k,t}$$

Here, mineral $i$ refers to eight battery minerals (lithium, cobalt, nickel, manganese, graphite, aluminum, copper, and phosphate); battery chemistry $j$ encompasses six different battery chemistries (NMC111, NMC523, NMC622, NMC811, NCA, and LFP); EV sales scenario $k$ refers to the low, medium, or high sales scenario; and time $t$ ranges from 2024 to 2032.

Then, the total mineral demand for mineral $i$ for EV sales scenario $k$ at time $t$ is:

$$\sum_j Mineral\ Demand_{i,j,k,t}$$

where $Mineral\ Demand_{i,j,k,t}$ is defined in the previous equation. This value gives us the total mineral demand for a specific mineral $i$ and low/medium/high sales scenario $k$ for a given year $t$.

Mineral Demand projections for the low, medium, and high sales scenario are shown in Tables 3.4-6 below.

|           | 2024   | 2025   | 2026   | 2027   | 2028   | 2029   | 2030   | 2031   | 2032    | Annual Average |
|-----------|--------|--------|--------|--------|--------|--------|--------|--------|---------|----------------|
| Lithium   | 7,698  | 7,654  | 7,697  | 7,666  | 7,595  | 7,569  | 7,517  | 7,482  | 47,886  | 12,085         |
| Cobalt    | 4,460  | 4,165  | 3,954  | 3,702  | 3,432  | 3,184  | 2,909  | 2,894  | 18,489  | 5,243          |
| Nickel    | 26,862 | 25,548 | 24,303 | 22,824 | 21,243 | 19,805 | 18,198 | 18,103 | 115,655 | 32,505         |
| Manganese | 3,823  | 3,500  | 3,324  | 3,113  | 2,886  | 2,676  | 2,443  | 2,431  | 15,529  | 4,414          |
| Graphite  | 70,587 | 71,238 | 72,514 | 73,087 | 73,252 | 73,823 | 74,177 | 73,833 | 472,625 | 117,237        |
| Aluminum  | 47,322 | 47,723 | 48,549 | 48,905 | 48,990 | 49,348 | 49,562 | 49,332 | 315,786 | 78,391         |
| Copper    | 27,784 | 28,088 | 28,609 | 28,851 | 28,930 | 29,167 | 29,316 | 29,181 | 186,791 | 46,302         |
| Phosphate | 14,115 | 15,062 | 16,133 | 17,047 | 17,856 | 18,753 | 19,646 | 19,561 | 125,317 | 29,277         |

Table 3.4:  Mineral demand projections (metric tons) – low sales scenario

|  | 2024 | 2025 | 2026 | 2027 | 2028 | 2029 | 2030 | 2031 | 2032 | Annual Average |
|---|---|---|---|---|---|---|---|---|---|---|
| Lithium | 15,395 | 20,411 | 25,656 | 30,664 | 34,063 | 37,616 | 41,005 | 44,440 | 47,886 | 33,015 |
| Cobalt | 8,919 | 11,107 | 13,179 | 14,807 | 15,392 | 15,825 | 15,869 | 17,189 | 18,489 | 14,531 |
| Nickel | 53,725 | 68,127 | 81,011 | 91,297 | 95,274 | 98,425 | 99,264 | 107,521 | 115,655 | 90,033 |
| Manganese | 7,646 | 9,334 | 11,080 | 12,451 | 12,941 | 13,300 | 13,328 | 14,437 | 15,529 | 12,227 |
| Graphite | 141,174 | 189,969 | 241,713 | 292,347 | 328,525 | 366,882 | 404,607 | 438,534 | 472,625 | 319,597 |
| Aluminum | 94,643 | 127,262 | 161,829 | 195,619 | 219,713 | 245,247 | 270,340 | 293,009 | 315,786 | 213,716 |
| Copper | 55,567 | 74,900 | 95,362 | 115,403 | 129,747 | 144,955 | 159,910 | 173,318 | 186,791 | 126,217 |
| Phosphate | 28,229 | 40,166 | 53,775 | 68,190 | 80,083 | 93,200 | 107,164 | 116,181 | 125,317 | 79,145 |

Table 3.5:  Mineral demand projections (metric tons) – medium sales scenario

|  | 2024 | 2025 | 2026 | 2027 | 2028 | 2029 | 2030 | 2031 | 2032 | Annual Average |
|---|---|---|---|---|---|---|---|---|---|---|
| Lithium | 49,453 | 49,174 | 49,447 | 49,249 | 48,794 | 48,626 | 48,295 | 48,068 | 47,886 | 48,777 |
| Cobalt | 28,651 | 26,759 | 25,400 | 23,782 | 22,048 | 20,456 | 18,690 | 18,592 | 18,489 | 22,541 |
| Nickel | 172,575 | 164,129 | 156,134 | 146,632 | 136,477 | 127,234 | 116,912 | 116,298 | 115,655 | 139,116 |
| Manganese | 24,561 | 22,486 | 21,354 | 19,998 | 18,538 | 17,193 | 15,698 | 15,616 | 15,529 | 18,997 |
| Graphite | 453,480 | 457,663 | 465,858 | 469,539 | 470,598 | 474,267 | 476,541 | 474,334 | 472,625 | 468,323 |
| Aluminum | 304,013 | 306,592 | 311,896 | 314,185 | 314,730 | 317,031 | 318,403 | 316,929 | 315,786 | 313,285 |
| Copper | 178,494 | 180,446 | 183,793 | 185,348 | 185,856 | 187,383 | 188,340 | 187,468 | 186,791 | 184,880 |
| Phosphate | 90,679 | 96,765 | 103,642 | 109,520 | 114,716 | 120,479 | 126,216 | 125,665 | 125,317 | 112,555 |

Table 3.6:  Mineral demand projections (metric tons) – high sales scenario

**SECTION IV: Sensitivity Tests**

We consider several sensitivity tests to assess the robustness of our results/underlying assumptions:

- An *Added Supply Assumption* where available production of each mineral increases by an amount equal to 20 percent of the annual production from the top producing country for that respective mineral. In the context of U.S. domestic content policies, such an increase could be interpreted in various ways: new production from free trade partners and domestic mine operators, loosened domestic content policies, establishment of free trade agreements with new international partners, boosted secondary production from recycling, or technological advances that increase the productivity of existing mines.

- A *Battery Pack Downsizing Assumption* where EV battery packs are downsized in capacity to hit the desired level of EV deployment in 2032 (6.2 million new EVs sold in the year 2032) under both current mineral production and the Added supply assumption.

- A *Heavier Fleet Assumption* in which true EV deployment is skewed towards a mix of 71 percent light trucks and SUVs and 29 percent sedans (consistent with existing fleet composition data), as opposed to our default case which considers a fleet of 100 percent sedans. We assume that light trucks and SUVs require a larger battery of approximately 100 kWh to achieve the target range of 300 miles, with correspondingly higher per-pack mineral requirements. We evaluate the potential ceiling to nationwide EV deployment under current and Added Supply mineral constraints for this heavier vehicle fleet.

***Part A: Added Supply Assumption***

We consider a scenario where the available production of each mineral increases by an amount equal to 20 percent of the annual production from the top producing country for that respective mineral. Estimates for annual supply under this scenario are presented in Table 4.1:

|  | Top producing country | Annual supply (tons) | Added supply (equal to 20 percent of top producing country's annual 2022 production) (tons) | Annual supply, added supply assumption (tons) |
|---|---|---|---|---|
| Lithium | Australia | 102,000 | 12,200 | 114,200 |
| Cobalt | Congo | 21,800 | 26,000 | 47,800 |
| Nickel | Indonesia | 933,000 | 320,000 | 1,253,000 |
| Manganese | South Africa | 3,530,000 | 1,440,000 | 4,970,000 |
| Graphite | China[a] | 48,000 | 170,000 | 218,000 |
| Aluminum | Australia | 26,524,000 | 5,000,000 | 31,524,000 |
| Copper | Chile | 11,289,000 | 1,040,000 | 12,329,000 |
| Phosphate | China | 84,950,000 | 17,000,000 | 101,950,000 |

Table 4.1: Annual supply in added supply assumption (metric tons)

[a] Only natural graphite is considered.

In this scenario, using a similar method as in Part B, we estimate the number of EV batteries that can be produced annually, with estimates shown in Table 4.2. As before, the limiting mineral for each battery chemistry is bolded. We find that NMC811 is the optimal battery chemistry, producing about 3.9 million EV batteries. However, this still falls short of the 5.7 million batteries needed to meet EPA targets in 2032.

|  | NMC111 | NMC523 | NMC622 | NMC811 | NCA | LFP |
|---|---|---|---|---|---|---|
| Lithium | 11,531,647 | 12,767,180 | 13,482,849 | 18,175,018 | 13,670,131 | 13,012,821 |
| Cobalt | **1,798,421** | 3,400,652 | **3,078,237** | 7,607,407 | 17,165,471 | n/a |
| Nickel | 47,949,566 | 35,020,308 | 27,737,538 | 25,566,143 | 20,928,612 | n/a |
| Manganese | 200,484,859 | 243,087,891 | 352,065,284 | 790,979,334 | n/a | n/a |
| Graphite | 3,629,824 | **3,218,900** | 3,088,541 | **3,854,985** | **3,558,456** | **2,258,235** |
| Aluminum | 704,855,665 | 704,855,665 | 676,696,910 | 836,178,153 | 754,706,140 | 489,829,862 |
| Copper | 482,419,414 | 482,419,414 | 459,664,608 | 490,542,465 | 520,878,882 | 324,198,589 |
| Phosphate | n/a | n/a | n/a | n/a | n/a | 2,788,071,049 |

Table 4.2. Number of EV batteries given each battery chemistry in the added supply assumption (annual production)

**Part B: Battery Pack Downsizing Assumption**

We further consider a thought experiment where EV battery packs are downsized in capacity to achieve the desired level of EV deployment in 2032 under both current mineral production and the Added Supply Assumption. This was estimated using the following equation:

$$MI_i = \frac{M_i}{EV^{Desired}}$$

where $MI_i$ refers to the requisite mineral intensity limit for mineral $i$ to satisfy the desired number of EVs; $M_i$ refers to current mineral supply for mineral $i$; and $EV^{Desired}$ refers to the desired number of EVs.

As previously calculated, the desired number of EVs in 2032 is 5.7 million. Estimates for the requisite mineral intensity, namely what mineral intensity would need to be to produce 5.7 million EVs, are shown in Table 4.3.

| | Requisite mineral intensity | |
|---|---|---|
| | Current mineral production | Added supply assumption |
| Lithium | 17.86 | 19.99 |
| Cobalt | 3.82 | 8.37 |
| Nickel | 163.35 | 219.37 |
| Manganese | 618.02 | 870.13 |
| Graphite | 8.40 | 38.17 |
| Aluminum | 4643.71 | 5519.09 |
| Copper | 1976.43 | 2158.51 |
| Phosphate | 14872.69 | 17848.98 |

Table 4.3: Requisite mineral intensity (kg/vehicle)

In both scenarios, graphite remains the limiting mineral in the production of batteries. Under current mineral production, the requisite intensity of graphite would be 8.4kg/vehicle. This suggests that EV battery capacities would have to shrink to the equivalent of about 56 percent of the size of a 20 kWh NMC 811 PHEV battery to produce the requisite number of EVs.

Under the Added Supply Assumption, the requisite intensity of graphite would be 38.2kg/vehicle. EV battery capacities would have to shrink to the equivalent of about 80 percent of the requirement of a 60 kWh NMC 811 battery to produce the requisite number of EVs.

*Part C: Heavier Fleet Assumption*

We consider a scenario in which true EV deployment is skewed towards a mix of 71 percent light trucks and 29 percent sedans, as opposed to our default case of 100 percent sedans. We assume that light trucks and SUVs require a larger battery of approximately 100 kWh to achieve the target range of 300 miles, with correspondingly higher per-pack mineral requirements. We evaluated the potential ceiling to nationwide EV deployment under current and Added Supply mineral constraints for this heavier vehicle fleet.

Table 4.4 presents a weighted average of the mineral content required for each battery chemistry based on this EV deployment mix.

|           | NMC 111 | NMC 523 | NMC 622 | NMC 811 | NCA   | LFP    |
|-----------|---------|---------|---------|---------|-------|--------|
| Lithium   | 12.25   | 11.06   | 10.47   | 7.77    | 10.33 | 10.85  |
| Cobalt    | 32.87   | 17.38   | 19.20   | 7.77    | 3.44  | 0.00   |
| Nickel    | 32.32   | 44.25   | 55.86   | 60.61   | 74.04 | 0.00   |
| Manganese | 30.66   | 25.28   | 17.46   | 7.77    | 0.00  | 0.00   |
| Graphite  | 74.27   | 83.75   | 87.29   | 69.93   | 75.76 | 119.38 |
| Aluminum  | 55.31   | 55.31   | 57.61   | 46.62   | 51.66 | 79.59  |
| Copper    | 31.60   | 31.60   | 33.17   | 31.08   | 29.27 | 47.03  |
| Phosphate | 0.00    | 0.00    | 0.00    | 0.00    | 0.00  | 45.22  |

Table 4.4: Weighted average of mineral content for battery chemistries (kg)

As before, we estimate the number of EV batteries that can be produced for each battery chemistry given current mineral production levels. Estimates are presented in Table 4.5 and Table 4.6 for current production levels and the added supply assumption. The limiting mineral for each battery chemistry is bolded.

Exclusive production of NMC811 allows cumulative deployment of approximately 4.12 million EVs under current production levels and about 18.70 million EVs under the Added Supply Assumption.

|  | NMC111 | NMC523 | NMC622 | NMC811 | NCA | LFP |
|---|---|---|---|---|---|---|
| Lithium | 8,328,626 | 9,220,979 | 9,737,837 | 13,126,721 | 9,873,113 | 9,398,390 |
| Cobalt | 663,236 | 1,254,119 | 1,135,213 | 2,805,515 | 6,330,407 | n/a |
| Nickel | 28,871,094 | 21,086,210 | 16,701,106 | 15,393,704 | 12,601,387 | n/a |
| Manganese | 115,145,744 | 139,614,215 | 202,203,317 | 454,287,512 | n/a | n/a |
| Graphite | **646,276** | **573,113** | **549,901** | **686,365** | **633,569** | **402,070** |
| Aluminum | 479,563,245 | 479,563,245 | 460,403,536 | 568,910,386 | 513,479,286 | 333,265,894 |
| Copper | 357,190,906 | 357,190,906 | 340,341,919 | 363,204,796 | 385,666,322 | 240,041,678 |
| Phosphate | n/a | n/a | n/a | n/a | n/a | 1,878,572,252 |

Table 4.5: Number of EV Batteries that can be produced for each battery chemistry (annual production)

|  | NMC111 | NMC523 | NMC622 | NMC811 | NCA | LFP |
|---|---|---|---|---|---|---|
| Lithium | 9,324,796 | 10,323,881 | 10,902,558 | 14,696,780 | 11,054,014 | 10,522,511 |
| Cobalt | **1,454,251** | 2,749,857 | **2,489,137** | 6,151,542 | 13,880,435 | n/a |
| Nickel | 38,773,292 | 28,318,351 | 22,429,245 | 20,673,431 | 16,923,406 | n/a |
| Manganese | 162,117,379 | 196,567,322 | 284,688,523 | 639,605,931 | n/a | n/a |
| Graphite | 2,935,172 | **2,602,888** | 2,497,469 | **3,117,239** | **2,877,458** | **1,826,069** |
| Aluminum | 569,965,003 | 569,965,003 | 547,193,525 | 676,154,841 | 610,274,507 | 396,089,355 |
| Copper | 390,097,146 | 390,097,146 | 371,695,944 | 396,665,066 | 421,195,862 | 262,155,536 |
| Phosphate | n/a | n/a | n/a | n/a | n/a | 2,254,507,841 |

Table 4.6: Number of EV Batteries that can be produced for each battery chemistry (annual production, Added Supply Assumption)

**SECTION V: Emissions Impact of Disequilibrium**

Based on the discrepancy between the demand and supply of EV batteries in the optimal mix scenario under current production levels, we determine the emissions impact of being unable to meet each EV sales volume target necessitated by the EPA proposal.

To do so, we leverage the Greenhouse Gases, Regulated Emissions, and Energy Use in Transportation (GREET) model commonly used in vehicle lifecycle emissions analyses in combination with a model developed in previous works (Woodley et al., 2023; Nunes et al., 2022; MIT Energy Initiative, 2019). We estimate per-mile emissions, considering vehicle manufacturing emissions, fuel usage and production emissions, fuel efficiency, aggregate utilization, and energy per gallon of gasoline using the following equation:

$$E_{PM} = \frac{((e_{vm} \times 1,000,000) + e_{vd} + e_{mr})}{au} + (\frac{1}{FE} * (\frac{e_{fp}}{MJ_E} + \frac{e_{fu}}{MJ_E}) * EC_g)$$

where $E_{PM}$ = emissions per mile (g CO2e/mi); $e_{vm}$ = vehicle manufacturing emissions (metric tons CO2-equivalent (CO2e)); $e_{vd}$ = emissions from end-of-life vehicle disposal; $e_{mr}$ = emissions from vehicle maintenance and repair; au = aggregate utilization (miles); FE = vehicle fuel efficiency (miles per gallon-equivalent (MPGe)); $\frac{e_{fp}}{MJ_E}$ = fuel production emissions ($g\ CO_2e$ per megajoule of energy); $\frac{e_{fu}}{MJ_E}$ = fuel usage emissions ($g\ CO_2e$ per megajoule of energy); and $EC_g$ = energy content of gasoline (lower heating value).

After estimating EV and ICEV per-mile emissions, we calculate lifecycle emissions using the following equation:

$$E_{PV} = \frac{au}{1,000,000} * E_{PM}$$

where $E_{PV}$ = emissions per vehicle (tons $CO_2e$); au = aggregate utilization (miles); and $E_{PM}$ = emissions per mile (g CO2e/mi).

We subtract lifecycle emissions for each HEV or EV battery chemistry from ICEV lifecycle emissions to find the emissions benefit relative to ICEVs. To illustrate, figures from 2023 estimates have been included in Table 5.1.

| | Total Per-Mile Emissions (g CO2e/mi) | Lifecycle Emissions (tons CO2e/vehicle) | Emissions Benefit Relative to ICEVs (tons CO2e/vehicle) |
|---|---|---|---|
| ICEV | 374.5 | 64.84 | - |
| HEV | 274.7 | 47.57 | 17.28 |
| EV NMC 111 | 158.4 | 27.43 | 37.41 |
| EV NMC 523 | 162.9 | 28.21 | 36.63 |
| EV NMC 622 | 163.5 | 28.31 | 36.53 |
| EV NMC 811 | 160.9 | 27.86 | 36.98 |
| EV NCA | 162.8 | 28.19 | 36.65 |

Table 5.1: 2023 estimates of emissions benefit relative to ICEVs

Using this methodology, we estimate lifecycle emissions for each battery chemistry from 2027 to 2032. These estimates are shown in Table 5.2. We account for improvements in fuel economy as well as improvements to the electric grid when calculating emissions in each year. Based on the US' target of a 50 percent emissions reduction (relative to 2005) by 2030 (13) – a goal further supported via the enactment of the Inflation Reduction Act (IRA) (14) –, we assume emissions associated with the electric grid decline linearly such that a 50 percent reduction relative to 2005 is achieved. Regarding ICEV and HEV fuel economy, we assumed an annual improvement rate of 8 percent through 2025 and 10 percent from 2026 – 2032, which is consistent with existing CAFE standards (15). However, owing to diminishing returns on further technical innovation, we impose a capped maximum fuel economy 75 miles per gallon for HEVs[2].

Next, we calculate the emissions shortfall from the shortage of EVs in each year, using the total number of EVs possible from annual production of minerals. We then calculate annual emissions shortfall in each sales scenario based on the following formula:

$$Emissions\ Shortfall_{k,t} = EV\ Batteries\ Shortfall_{k,t} \times Emissions\ Benefit_{k,t}$$

[2] PHEVs are excluded from our model given, 1) they offer fuel economy that is – on average – less advantageous than HEVs, 2) are more mineral intensive than HEVs to manufacture, and 3) consistently constitute less than 1 percent of light duty vehicle sales. We note that this approach is consistent with longstanding mineral supply analysis (16).

Here, $k$ refers to the sales scenario and $t$ refers to the year, $EV\ Batteries\ Shortfall$ refers to the possible number of EVs that may be produced, and $Emissions\ Benefit$ refers to the emissions benefit of the EV relative to ICEVs. As calculated in earlier sections, a maximum of 848,804 batteries may be produced in any given year using the NMC 811 battery chemistry.

Estimates of emissions shortfall by sales scenario and year are shown in Table 5.3.

| | 2027 | 2028 | 2029 | 2030 | 2031 | 2032 |
|---|---|---|---|---|---|---|
| ICEV Fuel Economy | 60.00 | 61.20 | 62.50 | 63.70 | 65.10 | 66.40 |
| HEV Fuel Economy | 73.39 | 75 | 75 | 75 | 75 | 75 |
| EV Fuel Economy | 114 | 114 | 114 | 114 | 114 | 114 |
| Electric Grid Emissions Rate (g $CO_2$e/kWh) | 309.78 | 304.12 | 298.56 | 293.10 | 287.74 | 282.48 |
| Lifecycle emissions – ICEV (tons $CO_2$e/vehicle) | 40.21 | 39.58 | 38.92 | 38.34 | 37.69 | 37.11 |
| Lifecycle emissions – HEV (tons $CO_2$e/vehicle) | 36.73 | 36.16 | 36.16 | 36.16 | 36.16 | 36.16 |
| Lifecycle emissions – EV NMC 111 (tons $CO_2$e/vehicle) | 26.34 | 26.08 | 25.83 | 25.58 | 25.34 | 25.09 |
| Lifecycle emissions – EV NMC 523 (tons $CO_2$e/vehicle) | 27.12 | 26.86 | 26.61 | 26.36 | 26.11 | 25.87 |
| Lifecycle emissions – EV NMC 622 (tons $CO_2$e/vehicle) | 27.22 | 26.97 | 26.71 | 26.46 | 26.22 | 25.98 |
| Lifecycle emissions – EV NMC 811 (tons $CO_2$e/vehicle) | 26.77 | 26.52 | 26.26 | 26.01 | 25.77 | 25.53 |
| Lifecycle emissions – EV NCA (tons $CO_2$e/vehicle) | 27.10 | 26.85 | 26.59 | 26.34 | 26.10 | 25.86 |
| Lifecycle emissions – EV LFP (tons $CO_2$e/vehicle) | 26.06 | 25.81 | 25.55 | 25.30 | 25.06 | 24.82 |

Table 5.2: Lifecycle emissions overview

| | | 2027 | 2028 | 2029 | 2030 | 2031 | 2032 |
|---|---|---|---|---|---|---|---|
| All sales scenarios | Projected light-duty vehicle sales | 15,478,700 | 15,330,200 | 15,268,900 | 15,210,400 | 15,144,000 | 15,102,000 |
| Low sales scenario | # of EVs desired | 911,257 | 902,515 | 898,906 | 895,462 | 891,553 | 5,711,810 |
| | # of EVs possible (Production) | 848,804 | 848,804 | 848,804 | 848,804 | 848,804 | 848,804 |
| | Optimal EV battery chemistry | NMC 811 | NMC 811 | NMC 811 | NMC 811 | NMC 811 | NMC 811 |
| | Emissions shortfall from lack of EVs (tons $CO_2$e) | 838,865 | 701,400 | 634,080 | 574,939 | 509,334 | 56,281,611 |
| Medium sales scenario | # of EVs desired | 3,645,234 | 4,047,671 | 4,467,348 | 4,884,425 | 5,295,399 | 5,711,810 |
| | # of EVs possible (Production) | 848,804 | 848,804 | 848,804 | 848,804 | 848,804 | 848,804 |
| | Optimal EV battery chemistry | NMC 811 | NMC 811 | NMC 811 | NMC 811 | NMC 811 | NMC 811 |
| | Emissions shortfall from lack of EVs (tons $CO_2$e) | 37,561,279 | 41,773,389 | 45,795,416 | 49,728,490 | 52,979,047 | 56,281,611 |
| High sales scenario | # of EVs desired | 5,854,284 | 5,798,119 | 5,774,935 | 5,752,809 | 5,727,696 | 5,711,810 |
| | # of EVs possible (Production) | 848,804 | 848,804 | 848,804 | 848,804 | 848,804 | 848,804 |
| | Optimal EV battery chemistry | NMC 811 | NMC 811 | NMC 811 | NMC 811 | NMC 811 | NMC 811 |
| | Emissions shortfall from lack of EVs (tons $CO_2$e) | 67,232,948 | 64,632,153 | 62,343,920 | 60,429,055 | 58,129,658 | 56,281,611 |

Table 5.3: EV Sales scenarios and emissions shortfall

**SECTION VI: Resolution Pathways**

***Part A:  Increasing Mineral Production Capacity***

For our first resolution pathway of increasing mineral production capacity, we estimate the production thresholds necessary to achieve the EPA's targets. Our model estimates that graphite, and to a lesser extent cobalt, are the minerals currently in greatest shortfall. Thus, increasing the production of both graphite and cobalt are most crucial to increasing the number of EV batteries that can be manufactured.

We calculate the mineral requirements to meet the EV deployment target of 5.7 million in the hypothetical market share scenario using the following formula:

$$Mineral\ Requirement_t = \frac{Target\ EV\ Deployment}{Max\ EV\ Deployment} \times Current\ Mineral\ Supply$$

where $\frac{Target\ EV\ Deployment}{Max\ EV\ Deployment}$ is the multiplying factor by which mineral production must increase by, calculated by dividing the target deployment (approximately 5.7 million) by the maximum EV deployment given existing mineral production.

Table 6.1 shows the minimum and maximum amount of graphite and cobalt needed. Figures in bold represent amounts that exceed current levels of production. Graphite production would need to increase by about 10 times from present day levels, and cobalt about 1.4 times from present day levels to meet mineral demand.

|  | Current Production | Minimum Desired Amount | Maximum Desired Amount |
|---|---|---|---|
| Graphite | 48,000 | **431,520** | **473,280** |
| Cobalt | 21,800 | 18,615 | **31,174** |

Table 6.1: Current production, minimum and maximum desired amounts of graphite and cobalt

Table 6.2 further shows potential future annual graphite production by the U.S. and partner countries, based on recent project announcements. Despite a projected increase in graphite production to 255,000 tons by 2032, this is still insufficient to meet the EV deployment target.

| | 2023 | 2024 | 2025 | 2026 | 2027 | 2028 | 2029 | 2030 | 2031 | 2032 |
|---|---|---|---|---|---|---|---|---|---|---|
| Annual graphite production (millions kg) | 58.0 | 69.25 | 109.25 | 173.0 | 173.0 | 255.0 | 255.0 | 255.0 | 255.0 | 255.0 |

Table 6.2: Potential future annual graphite production by the U.S. and partner countries based on recent project announcements (millions kg)

We next consider the number of EVs that can be produced under the Added Supply Assumption. The number of EVs that can be produced annually given projected increases in mineral production levels is shown in Table 6.3. Cumulatively, about 23.13 million EVs can be deployed in the optimal chemistry scenario if NMC811 was exclusively relied upon and 15.95 million EVs in the market mix scenario.

| | 2027 | 2028 | 2029 | 2030 | 2031 | 2032 |
|---|---|---|---|---|---|---|
| Optimal Chemistry Scenario | 3,854,985 | 3,854,985 | 3,854,985 | 3,854,985 | 3,854,985 | 3,854,985 |
| Market Mix Scenario | 2,717,719 | 2,685,716 | 2,651,851 | 2,629,745 | 2,631,540 | 2,631,540 |

Table 6.3: Number of EV batteries that can be produced under the Added Supply Assumption

Additionally, we consider a Battery Pack Downsizing Assumption to stretch mineral supplies further. As found in the earlier sensitivity tests, at present-day rates of mineral production, battery capacity would need to decrease to around 11 kWh. The Added Supply Assumption allows for up to 38.2kg/vehicle, hence battery capacity would need to decrease to 48 kWh.

Finally, we recognize that our analysis optimistically assumes battery packs scaled for sedan-sized electric vehicles. Hence, we also considered a Heavier Fleet Assumption where 71 percent of new EVs are SUVs and pickup trucks carry larger batteries to achieve 300-mile range. Table 6.4 shows the number of EVs that can be produced given projected increases in mineral production levels and a heavier fleet for the optimal chemistry and market mix scenarios. Cumulatively, approximately 18.70 million EVs can be deployed in the optimal chemistry scenario if NMC811 was exclusively produced, and 12.90 million in EVs in the market mix scenario.

|  | 2027 | 2028 | 2029 | 2030 | 2031 | 2032 |
|---|---|---|---|---|---|---|
| Optimal Chemistry Scenario | 3,117,239 | 3,117,239 | 3,117,239 | 3,117,239 | 3,117,239 | 3,117,239 |
| Market Mix Scenario | 2,197,618 | 2,171,740 | 2,144,356 | 2,126,481 | 2,127,932 | 2,127,932 |

Table 6.4: Number of EV batteries that can be produced under the Heavier Fleet Assumption and Added Supply Assumption

***Part B: Using HEVs to Realize Emissions Reductions***

In our second pathway, we consider using HEVs to achieve the emissions benefit envisioned by the EPA proposal instead of NMC 811 EVs.

We first calculate the emissions benefit from replacing an ICEV with a HEV in each year. By subtracting the expected total lifetime emissions of an ICEV from that of a HEV, we calculate the emissions benefit from replacing an ICEV with the following types of vehicles. The results are shown in Table 6.5.

|        | 2023  | 2024  | 2025  | 2026  | 2027  | 2028  | 2029  | 2030  | 2031  | 2032  |
|--------|-------|-------|-------|-------|-------|-------|-------|-------|-------|-------|
| NMC111 | 37.41 | 33.48 | 29.86 | 25.69 | 13.86 | 13.49 | 13.09 | 12.76 | 12.35 | 12.01 |
| NMC523 | 36.63 | 32.70 | 29.08 | 24.91 | 13.09 | 12.71 | 12.31 | 11.98 | 11.57 | 11.23 |
| NMC622 | 36.53 | 32.60 | 28.97 | 24.81 | 12.98 | 12.61 | 12.21 | 11.87 | 11.46 | 11.12 |
| NMC811 | 36.98 | 33.05 | 29.42 | 25.26 | 13.43 | 13.06 | 12.66 | 12.32 | 11.91 | 11.57 |
| NCA    | 36.65 | 32.72 | 29.09 | 24.93 | 13.10 | 12.73 | 12.33 | 11.99 | 11.59 | 11.24 |
| LFP    | 37.69 | 33.76 | 30.13 | 25.97 | 14.14 | 13.77 | 13.37 | 13.03 | 12.62 | 12.28 |
| HEV    | 17.28 | 15.82 | 14.47 | 12.94 | 3.48  | 3.41  | 2.75  | 2.17  | 1.52  | 0.94  |

Table 6.5: Emissions benefit from replacing an ICEV with alternative powertrain/chemistries (tons $CO_2e$)

We calculate the number of HEVs needed to meet emissions reduction targets using the following formula:

$$HEV_t^{Desired} = \frac{EB_t^{Desired}}{EB_t^{HEV}}$$

where $HEV_t^{Desired}$ refers to number of HEVs needed; $EB_t^{Desired}$ refers to the administration's desired emissions benefit; and $EB_t^{EV}$ refers to the emissions benefit.

Table 6.6 shows the desired number of HEVs in each sales scenario as well as projected light-duty vehicle sales in the US. Figures which exceed the projected light-duty vehicle sales are bolded. In all sales scenarios, the desired number of HEVs eventually exceeds projected light-duty vehicle sales.

Hence, we next considered whether HEVs can supplement EVs to achieve emissions targets. We estimate the minimum number of EVs required such that, given that HEVs constitute the remainder of light vehicle sales, the total emissions benefit is equal to the desired emissions benefit. This was calculated using the following equation:

$$\min_{EV} \ (EV_t^{Hypothetical} \times EB_t^{EV}) + [(LD_t - EV_t^{Hypothetical}) \times EB_t^{HEV}] = Total_i$$

$$s.t. Total_t = EB_t^{Desired}$$

where $EV_t^{Hypothetical}$ refers to the EV sales required to supplement HEVs in the hypothetical sales scenario in year t; $EB_t^{EV}$ refers to the emissions benefit from electric vehicles in year t; $LD_t$ refers to light-duty sales in year t; and $EB_t^{Desired}$ refers to the administration's desired emissions benefit in year t.

Figures are shown in Table 6.7. In the medium sales scenario, about 2.7 million EVs are required in 2030, and in the high sales scenario about 2.5 million EVs are required in 2027. In all scenarios, 4.9 million EVs are required in 2032.

| | 2027 | 2028 | 2029 | 2030 | 2031 | 2032 | Total |
|---|---|---|---|---|---|---|---|
| US Projected Light-Duty Vehicle Sales | 15,478,700 | 15,330,200 | 15,268,900 | 15,210,400 | 15,144,000 | 15,102,000 | 91,534,200 |
| Desired HEVs in Low Sales Scenario | 3,520,491 | 3,455,756 | 4,131,384 | 5,082,267 | 6,994,580 | **70,516,195** | **93,700,673** |
| Desired HEVs in Medium Sales Scenario | 14,082,754 | **15,498,650** | **20,531,987** | **27,721,949** | **41,544,465** | **70,516,195** | **189,896,001** |
| Desired HEVs in High Sales Scenario | **22,617,051** | **22,201,166** | **26,541,673** | **32,650,533** | **44,936,004** | **70,516,195** | **219,462,622** |

Table 6.6: Comparison of desired HEVs in different sales scenarios and US projected light-duty vehicle sales

| | 2027 | 2028 | 2029 | 2030 | 2031 | 2032 |
|---|---|---|---|---|---|---|
| Minimum EVs in Low Sales Scenario | 0 | 0 | 0 | 0 | 0 | 4,905,031 |
| Minimum EVs in Medium Sales Scenario | 0 | 164,616 | 1,552,892 | 2,736,122 | 3,893,587 | 4,905,031 |
| Minimum EVs in High Sales Scenario | 2,569,650 | 2,517,508 | 3,213,290 | 3,785,172 | 4,387,415 | 4,905,031 |

Table 6.7: Minimum number of EVs required to supplement HEVs